%% file: pap1.tex
\documentclass[seceqn]{elsart}
\usepackage{macros}
\usepackage{macros_static}
\usepackage[dvips]{epsfig}
\begin{document}
\input title.tex
\input sect1.tex
\input sect2.tex
\input sect3.tex

\input sect4.tex
\input sect5.tex

\begin{appendix}
\input appa.tex
\input appb.tex

\input appc.tex

\end{appendix}
\bibliography{pap1}      
\bibliographystyle{h-elsevier}    
\end{document}

%% file: title.tex
\begin{frontmatter}

\begin{flushright}
DESY 00-091 \\
hep-lat/0007002
\end{flushright}

\title{
Renormalization and O($a$)-improvement of the static axial current
}
\vbox{
\centerline{
\epsfxsize=2.5 true cm
\epsfbox{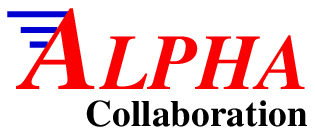}}
}
\author[DESY]{Martin Kurth}
\author[DESY]{Rainer Sommer}
\address[DESY]{
Deutsches Elektronen-Synchrotron, DESY \\
Platanenallee 6, D-15738 Zeuthen, Germany
}

\maketitle
\begin{abstract}

A systematic treatment of $\Oa$--improvement
in lattice theories with static quarks is presented.  
The Schr\"odinger functional is discussed
and a renormalization condition for the static axial current
in the SF-scheme is introduced. 
Its relation to other schemes is computed to
1-loop order 
and the 2-loop anomalous dimension is derived.
In finite volume renormalization schemes such as the SF-scheme,
the renormalization scale dependence of the renormalized quantities
is described by the step scaling function which 
can be computed by MC-simulations. We evaluate its  
lattice spacing effects  in perturbation theory. 

\end{abstract}

\keyword{
lattice QCD; heavy quark effective theory; $\Oa$ improvement; 
renormalization;}\\
\PACS{11.15.Ha; 12.15.Hh; 12.38.Bx; 12.38.Gc; 14.65.Fy}
\endkeyword

\end{frontmatter}
\cleardoublepage

%% file: sect1.tex
\section{Introduction   \label{s_intro}}

CP-violation
in the B-system is the subject of an active experimental
programme
\cite{BaBar:Report,BELLE:Report,HERAB:Report}.
The interpretation of experiments in the framework of
the Standard Model of particle physics or possible extensions
requires substantial theoretical input, in particular
some knowledge of hadronic matrix elements. One important piece
in this puzzle is
the  leptonic decay constant of the B-meson, $\Fb$,
which in principle can be computed 
accurately by means of lattice QCD. 

Due to recent progress
in this field \cite{reviews:leshouches,reviews:tampere}, a
precise computation of 
decay constants of light mesons (e.g. $\Fk$) is
rather straight forward, at least in the quenched approximation
\cite{mbar:pap3}. The primary reason is that the renormalization of
the axial current and the improvement of action and current 
are known 
non-perturbatively~\cite{impr:pap1,impr:pap2,impr:pap3,impr:pap4}.
Heavy quarks are more difficult, since their mass is large
in comparison to the confinement scale and consequently
lattices with a 
large number of points are necessary to control lattice spacing 
effects (see e.g. \cite{reviews:beauty,reviews:wittig97} 
for reviews of this problem).
As an alternative to the standard treatment of quarks
as relativistic Dirac fields, approximate treatments, such as 
non-relativistic lattice QCD,
have therefore been developed to treat heavy quarks.\footnote{ 
For a  recent review of heavy quark physics on the lattice, 
we refer the reader to \cite{lat99:hashimoto}.}

On the other hand, in the limit $m_{\rm quark}\to\infty$,
the theoretical treatment simplifies, as 
in this limit quarks can be described by the ``static effective 
field theory'', which is believed to be
a local renormalizable quantum field 
theory \cite{stat:eichhill1,Neubert:1994mb}. 
Besides the theoretical interest
in this limit of QCD, where new symmetries
appear~\cite{stat:symm1,stat:symm2,stat:symm3},
there are relevant phenomenological applications.
Computations of quantities such as
$\Fb$ in the static limit provide 
important checks on the systematic errors made when
approximate methods are used. Furthermore, a very useful
strategy for the computation of B-meson matrix elements might
be an {\em interpolation} between the static limit and 
quark masses of order 1 GeV rather than an extrapolation 
from light quarks to heavy ones. This approach has been 
tried in the past in order to compute $\Fb$
\cite{FB:interpol1,reviews:beauty,reviews:wittig97},
but was limited in precision due to two reasons. First,
the renormalization of the current in
the static effective 
field theory was known only perturbatively
and second the computation of the matrix element in that limit
turned out to be rather difficult. In this paper we
assume that the second difficulty can be resolved by the methods
of \cite{mbar:pap2} or \cite{michael:alltoall} possibly combined
with the ideas of \cite{stat:fnal1}. 
For the time being, we concentrate on
preparing the ground for removing the first obstacle.
To this end we introduce a renormalization condition in the 
\SF as an intermediate renormalization
scheme. The setup is such that this scheme 
can be used for a non-perturbative computation of the
current renormalization necessary to obtain the
renormalization group invariant and scheme independent current,
following the strategy described in
\cite{lat99:zastat}.  

A previous attempt to renormalize the static axial current 
non-per\-tur\-ba\-tive\-ly \cite{zastat:MaSa} failed
when it was applied numerically  \cite{zastat:MaSapriv}. We have checked that
our renormalization condition yields good precision when evaluated 
in Monte Carlo computations \cite{zastat:nonpert}. 

Since a systematic discussion of $\Oa$ improvement of 
action and axial current 
in the static effective field theory has not been carried out before,
we present it in \sect{s_lattice}. This discussion has some relevance also in
the context of non-relativistic QCD. \\
In \sect{s_SF} we introduce the \SF for static quarks, discussing the 
continuum formulation and the lattice formulation as well as the 
perturbative expansion of certain correlation functions.\\
In \sect{s_renorm} we then describe the renormalization of the static axial
current in various renormalization schemes, introducing in particular
the SF-scheme which is suitable for non-perturbative MC computations. \\
The two-loop anomalous dimension of the current in the SF-scheme
is presented in \sect{s_gamma} and lattice spacing effects in the SF-scheme
are computed to one-loop order in \sect{s_ssf}. After the  one-loop 
computation of
$\castat$ and $\bastat$, coefficients necessary for $\Oa$-improvement of
the axial current, we finish with a 
detailed summary of our results.  Appendices contain our notations,
some details of the perturbative computations and tables with 
numerical results for certain coefficients in the perturbative expansions.

%% file: sect2.tex
\section{Static quarks on the lattice \label{s_lattice}}
In  this section we discuss QCD with a number $\nf$ of light quarks,
which for simplicity are taken to be mass-degenerate,
and one additional heavy quark flavour treated in the \stat. 
Generalizations are straight forward.
The theory is formulated
on infinite Euclidean space-time, approximated by  a
hypercubic lattice with spacing $a$. All statements
and formulae are unchanged if one considers
the theory on a torus, imposing suitable periodic boundary
conditions.

In order to present a coherent discussion, we start from the
well-known arguments about renormalization
of the effective theory and then proceed to analyse
what is necessary to achieve $\Oa$ improvement.

\subsection{Basic formulation\label{s_basic}}
The \stat is defined by a path integral over Grassmann-valued fields
$\heavy$ and $\heavyb$ which are associated to the lattice sites. 
They carry colour and Dirac indices.
For ease of notation they are taken as
4-component spinors, but satisfy the constraint
\be
  \label{e_constraint}
  P_{+}\heavy=\heavy\,,\quad \heavyb P_{+}=\heavyb\,,\quad 
  P_{+}=\frac12(1+\gamma_0)\,,
\ee
leaving us with two degrees of freedom per space-time point 
(the complementary components 
are set to zero). We follow~\cite{stat:eichhill1} and choose
\be
  \label{e_LatLag}
  \Sstat=a^4 \frac{1}{1+a\dmstat} 
            \sum_x \heavyb(x)(\nabstar{0} + \dmstat) \heavy(x)\,,
\ee
as our lattice action.  The definition of 
the backward lattice derivative, $\nabstar{\mu}$, and  
other conventions are summarized in \app{a_conv}. 
The factor
$1/ (1+a\dmstat)$ is convenient, but irrelevant for
physics, since it may be absorbed into the normalization
of the static fields.
The above Lagrangian describes quarks,
which, being treated in the static approximation, propagate only
forward in time. 

For completeness we also give the
action describing the propagation of static anti-quark fields, 
$\aheavy$, $\aheavyb$:
\be
 \Sastat = a^4 \frac{1}{1+a\dmstat}
               \sum_x \aheavyb(x)(\nab{0} + \dmstat) \aheavy(x)\,
\ee
with
\be
  P_{-}\aheavy=\aheavy\,,\quad \aheavyb P_{-}=\aheavyb\,,\quad 
  P_{-}=\frac12(1-\gamma_0)\,.
\ee 
Since the properties of  $\aheavy$, $\aheavyb$ are completely
analogous to the ones of $\heavy$, $\heavyb$ we restrict ourselves
to  the latter from now on. The dynamics of
gauge fields and relativistic quarks is governed by the 
action $\Sg+\Sf$ \cite{impr:sw,impr:pap1} 
listed in \app{a_conv}.

The time component of the
axial current composed of a static quark and a relativistic 
anti-quark field, $\lightb$, assumes a form,
\be
  \label{e_StatAxial}
  \Astat(x)=\lightb(x)\gamma_0 \gamma_5\heavy(x)\,,
\ee
familiar from the relativistic case. 
Observables are defined in the usual way through a path integral
with total action
\be
 S = \Sf+\Sg +\Sstat \,.
\ee

\subsection{Extra symmetries\label{s_symm}}

In our discussion of renormalization and $\Oa$-improvement 
we will need two symmetries which are present in the effective field theory,
but not in finite mass QCD. \\[1ex] 
1) The static action, \eq{e_LatLag}, (as well
as the integration measure) is invariant
under the $\SUtwo$ rotations,
\bes
  \label{e_spin}
  && \heavy\longrightarrow V\heavy,
  \qquad \heavyb\longrightarrow\heavyb V^{-1}\,,\\
{\rm with} &&\nonumber \\
  && V=\exp(-i\phi_i \epsilon_{ijk}\sigma_{jk})\,, 
\ees
and transformation parameters $\phi_i$.
This property is called {\em heavy quark spin symmetry}
\cite{stat:symm1,stat:symm2,stat:symm3}. \\[1ex]
2) Static quarks propagate only in time. The heavy quark flavour
number is therefore conserved {\em locally}. Mathematically this may
be formulated as the invariance of action and measure under
\be
  \label{e_quarknumber}
   \heavy\longrightarrow \rme^{i\eta(\vecx)}\,\heavy,
  \qquad \heavyb\longrightarrow\heavyb \rme^{-i\eta(\vecx)}, 
\ee
with an arbitrary space (but not time) dependent parameter $\eta(\vecx)$.\footnote{
We thank U. Wolff for pointing this invariance out to us.}
The axial current carries a flavour number corresponding to the 
transformation property 
\be
   \label{e_qn_current}
   \Astat(x) \longrightarrow \rme^{i\eta(\vecx)}\,\Astat(x)\,.
\ee
As a consequence of the invariance of the action
under \eq{e_quarknumber},
the propagator of the static quarks is proportional
to a (lattice) $\delta$-function in space (see \eq{e_prop}, below).
We may say that the quarks are ``exactly static'' 
for finite lattice spacing.
  
\subsection{Renormalization
            \label{s_subtr}}
Renormalization of the effective theory has been discussed before
\cite{Neubert:1994mb},
also in the context of the lattice regularized theory
\cite{stat:boucaud_za,stat:eichhill_za,stat:boucaud_rig,Maiani:1992az}.
Using the above symmetries, the discussion is quite simple 
and may be elevated also to the level of $\Oa$-improvement as will be done below.

We point out, however, that one essential assumption has to be made.
We assume that both Lagrangian and local composite fields can be made
finite by addition of local counterterms of same or lower canonical dimension.
This has been shown to be true to all orders of perturbation
theory for the relativistic theory \cite{Reisz:1989kk} but 
not for the effective theory. Of course 
this is the assumption  
that the static limit is
described by a {\em local} effective field theory. Perturbative computations 
performed in the past
\cite{Neubert:1994mb,stat:boucaud_za,stat:eichhill_za,stat:boucaud_rig,BorrPitt} and
also the ones presented below, provide non-trivial checks that
this assumption is justified.

With this assumption, we note simply that
\eq{e_LatLag}
contains all local terms with mass dimension $d \leq 4$ which comply with the 
standard (lattice-) rotational symmetry in the three space dimensions
as well as the
heavy quark spin symmetry.\footnote{
The additional flavour symmetry, which is present when
several flavours of static quarks (e.g. b and c) are considered
\cite{stat:symm1,stat:symm2,stat:symm3},
does not play a r\^ole in our context.} 
Apart from 
$\dmstat$, no counterterm is needed. 

At first sight, the presence of $\dmstat$ is worrying,
as -- for dimensional reasons -- it is linearly divergent, $\dmstat \propto 1/a$.
Unlike the corresponding mass counter term in the 
relativistic theory, there is no symmetry which allows a
non-perturbative determination of $\dmstat$.
Indeed, this represents a considerable problem in the determination
of the b-quark mass from lattice results in the static approximation 
\cite{Martinelli:1998vt}. It is therefore important to realize that
the dependence of correlation functions on $\dmstat$ is known 
explicitly and one can easily find combinations where 
$\dmstat$ cancels out.

As an example, we consider the correlation function
\be
  \label{e_caa}
    \caa(x_0) =  \sum_{\vecx} 
          \langle \Astat(x)  (\Astat)^{\dagger}(0) \rangle
\ee
with $ (\Astat)^{\dagger}(x)=\heavyb(x)\gamma_0\gamma_5\light(x)$.
Performing the integration over the quark fields (static and 
relativistic) yields
\be 
   \caa(x_0) = \sum_{\vecx} 
    \langle \tr S_{\rm h}(x,0) S^\dagger(x,0)\rangle_{\rm G}  
\ee
with a static propagator,
\bes
 \label{e_prop}    
  S_{\rm h}(x,y)&=& \theta(x_0-y_0)\, \delta(\vecx-\vecy)\nonumber\\
   & & \qquad
\times(1+a\dmstat)^{-(x_0-y_0)/a} \,  W(y,x)^{-1} P_{+}\nonumber \\[1.5ex] 
         W(x,x)&=&1    \\[1.5ex] 
         W(x,x+R\hat{\mu})&=& U(x,\mu) U(x+a\hat{\mu},\mu)
 \times\ldots\nonumber\\
 & &\qquad
 \times  U(x+(R-a)\hat{\mu},\mu) \nonumber
              \;\hbox{ for }\;R>0 \nonumber 
\ees
and a relativistic propagator, $S(x,y)$, satisfying the Dirac equation\\
$(D + m_0)S(x,y) = \delta(x-y)$. 
Here $D$ denotes the Dirac operator,
\eq{e_dirop}. The gauge field average $\langle . \rangle_{\rm G}$ 
is to be performed
with action $\Sg-\log(\det (D+m_0))$ and the trace extends over Dirac and 
colour 
indices. The functions $\delta,\theta$ are defined in 
eqs.(\ref{e_delta},\ref{e_theta}). 

If one starts with $\dmstat=0$ (and after renormalizing
the mass of the relativistic quarks), 
one discovers that 
the correlation function \eq{e_caa} is linearly divergent. This 
divergence is due to the well known self energy of a static quark,
\cite{stat:eichhill_za}
\be
 E_{\rm self} = a^{-1} \left\{ e_1 g_0^2 + \rmO(g_0^4) \right\}\,,
  \quad e_1={1\over{12\pi^2}}\times 19.95.
\ee
It has to be cancelled by a proper choice of $\dmstat$ to obtain a 
finite continuum limit of correlation functions
in the {\stat}.  The finite part of $\dmstat$ is a matter
of convention (scheme) as usual. Just to give an example, one could
require 
\be
 \label{e_fixdm}
 (\drv{0}+\drvstar{0})\log(\caa)(x_0) = 0\quad {\rm at} \quad x_0=1/\mu
\ee
for some renormalization scale $\mu$ as a condition to fix  $\dmstat$. 
With such a convention
\be
  (\caa)_\rmR(x_0) = (\zastat)^2 \caa(x_0) 
\ee
is finite for a suitably chosen renormalization constant
$\zastat$, since there is no other local composite field with dimension
$d\leq3$ and the quantum numbers of $\Astat$ which could induce mixing. 

Alternatively we may use the fact that due to \eq{e_prop}
\be
 \caa(x_0) = \caa(x_0)|_{\dmstat=0} \times (1+a\dmstat)^{-x_0/a}
\ee
and
form ratios of
correlation functions where $\dmstat$ cancels. One then avoids 
a  normalization condition such as \eq{e_fixdm}.
This is quite convenient as also for a determination
of $\Fb$ no knowledge of $\dmstat$ is required: it may be determined from
$(\caa)_\rmR(x_0)  \sim_{x_0\to\infty}  \Fb^2 \mb \exp(-E x_0)$,
where $E$ but not $\Fb$ depends on $\dmstat$.
We shall adopt this procedure in our definition of $\zastat$ in
the \SF in \sect{s_SF}.

\subsection{$\Oa$ improvement \label{s_Oa}}

We now discuss, which terms have to be added to action
and current, in order to have a lattice formulation
where (renormalized) on-shell observables approach the continuum
with a rate proportional to $a^2$. Such a
formulation is called $\Oa$-improved, as the leading 
lattice spacing corrections proportional to $a$ are 
removed.  
We follow the argumentation of \cite{impr:pap1}, where
$\Oa$-improvement is discussed for the theory with relativistic
quarks. We assume that the action $\Simpr$, \eq{e_simpr}, \cite{impr:sw,impr:pap1} 
is chosen
for these fields and consider only the modifications
necessary in $\Sstat$ and the static currents. 

\subsubsection{Action}
According to Symanzik, the cutoff effects of a lattice theory 
can be described in terms
of an effective {\em continuum} theory. 
One starts \cite{impr:pap1} from the terms which 
contribute to Symanzik's effective action~\cite{impr:Sym2}.
These may then be cancelled by adding terms to the lattice action,
resulting in an improved action. 

Symanzik's  action contains the lattice spacing as an explicit parameter.
It is written as
\be
  S_{\rm eff} = S_0 + a S_1 
\ee
to the order in the lattice spacing considered. $S_0$ is the standard 
continuum action and $S_1=\int\,\rmd^4x\lag{1}(x)$,
with $\lag{1}$ a linear combination
of local fields of dimension five 
with the appropriate symmetries~\cite{impr:pap1}. 
A basis of such fields is
given by
\begin{eqnarray}
  \label{e_basis}
  \op1 & = & \heavyb D_0D_0\heavy+\heavyb\ola{D}_0\ola{D}_0\heavy, 
   \nonumber \\
  \op2 & = &  m (\heavyb D_0 \heavy-\heavyb\ola{D}_0\heavy), \\
  \op3 & = &  m^2 \heavyb \heavy,  \nonumber 
\end{eqnarray}
where $D_\mu$ denotes the covariant derivative.
The mass-parameter, $m$, is the mass of the relativistic quarks.
There is no mass in the effective theory 
($\dmstat$ is a counter term), which could appear here.
Furthermore, terms $\heavyb\sigma_{0j}F_{0j}\heavy$ and
$\heavyb\sigma_{jk}F_{jk}\heavy$ 
do not appear
since $P_{+}\sigma_{0j} P_{+}=0$ and since the latter term
is not spin-symmetry invariant.
The reader might also miss a term
\be
 \heavyb D_jD_j\heavy+\heavyb\ola{D}_j\ola{D}_j\heavy
\ee
in our list. It is not invariant under the 
symmetry transformation \eq{e_quarknumber}.

The set of operators, \eq{e_basis}, can further be 
reduced~\cite{impr:onshell,impr:pap1} 
by using the formal equations of motion,
\be
   \label{e_eom}
   D_0\heavy(x)=0\,,\qquad \heavyb(x)\ola{D}_0=0\,.
\ee
They yield immediately
that both $\op1$ and $\op2$ vanish on-shell. 
The field $\op3$ contributes a mass-dependence 
to $\dmstat$. Such terms play a r\^ole in mass-independent renormalization 
schemes \cite{impr:pap1}.

In order to cancel the $\Oa$-effects, 
it suffices to modify $\dmstat$ or
equivalently to add a
correction term
\be
  \delta S_{\rm h} = a^5 \sum_x \,\bstat \, \mq^2 \, 
   \heavyb(x)  \heavy(x) \,,
\ee
to the lattice action. The coefficient $\bstat$ is a function of the bare 
coupling $g_0$, only; 
$\mq$ denotes the subtracted bare quark mass \cite{impr:pap1}. 
Obviously $\bstat$ vanishes in the quenched approximation
and is of order $\rmO(g_0^4)$ in general.

As explained at the end of
\sect{s_subtr}, for many applications one may consider
observables where $\dmstat$ and therefore also $\bstat$ drop out.

\subsubsection{Axial current}

Similarly to the action, renormalized composite lattice fields 
are described by effective fields in Symanzik's theory. For the
time component of the axial current we find
\be
 (\Astat)_{\rm eff} = \Astat + a \sum_{k=1}^4 \omega_k (\delta\Astat)_k
\ee
with some coefficients $\omega_k$. A basis for 
$\left\{(\delta\Astat)_k\right\}$ is  
\begin{eqnarray}
  \label{e_cur_basis}
  (\delta\Astat)_1 & = & \lightb\ola{D}_j\gamma_j\gamma_5\heavy,  
  \nonumber\\
  (\delta\Astat)_2 & = & \lightb\gamma_5D_0\heavy, \\
  (\delta\Astat)_3 & = & \lightb\ola{D}_0\gamma_5\heavy,  
  \nonumber\\
  (\delta\Astat)_4 & = & m\lightb\gamma_0\gamma_5\heavy. 
  \nonumber
\end{eqnarray}
Terms containing $D_j\heavy$ do not need to be taken into account 
since they violate \eq{e_qn_current}.
Requiring improvement to hold only on-shell, this set can be reduced:
From \eq{e_eom} one concludes that $(\delta\Astat)_2$ vanishes.
Due to $\heavy=\gamma_0\heavy$, 
the operators $(\delta\Astat)_1$, $(\delta\Astat)_3$ and
$(\delta\Astat)_4$ are connected by the (Dirac) field equation for
the relativistic quark. This means that one of these terms can be
eliminated. We keep  
$(\delta\Astat)_1$ and $(\delta\Astat)_4$. 

In order to achieve 
a cancellation of the $\Oa$ lattice spacing effects, we
add a corresponding combination of correction terms to the
axial current in the lattice theory. 
We then write the improved and renormalized current in the form
\bes
 \Aren &=& \zastat\,(1+\bastat a\mq) \Astatimpr \\
   \Astatimpr  &=&  \Astat +  a\castat \lightb\gamma_j\gamma_5
       \frac12(\lnab{j}+\lnabstar{j})\heavy
   \,,
\ees
with a mass-independent renormalization constant $\zastat$ and
improvement coefficients, $\bastat,\castat$, depending again 
on $g_0$ but not on $\mq$.

\subsubsection{Vector current}

The previous discussion carries over to the time component of the vector current.
Its improved and renormalized version may be parameterized as
\bes
 \Vren &=& \zvstat\,(1+\bvstat a\mq) \Vstatimpr \\
   \Vstatimpr  &=&  \lightb \gamma_0 \heavy  +  
           a\cvstat \lightb\gamma_j \frac12(\lnab{j}+\lnabstar{j})\heavy
   \,.
\ees
The space components of the vector (axial) current are related by the spin symmetry 
to the time component of the axial (vector) current. Therefore they do not need a 
separate discussion. 

\subsubsection{Relation to non-relativistic QCD \label{s_NRQCD}}

The \stat may be considered to be the infinite mass limit
of non-relativistic QCD (NRQCD).  Also in
NRQCD
the action is expanded in terms of composite fields of increasing
dimension. Conventionally the coefficients are assumed
to be proportional to inverse powers of the quark mass times
``logarithms''. When one wants to formulate  non-relativistic QCD
including $\Oa$-improvement, additional terms proportional
to the lattice spacing are needed. These are just the terms
discussed above. 

The necessity of a term proportional to $\castat$ has first been noted 
by Morningstar and Shigemitsu~\cite{MORNINGSTAR1}
in a 1-loop calculation in lattice NRQCD. 
They quote
\be
  c^{\rm stat}_{\rm A}={c^{\rm stat}_{\rm A}}^{(1)}g_0^2+{\rm O}(g_0^4),
\ee
with
\be
  \label{castat}
  {c^{\rm stat}_{\rm A}}^{(1)}=-{1 \over 4\pi}\times 1.00(1).
\ee
Our analysis implies that additional terms 
proportional to the mass of the relativistic quarks
are necessary to achieve full $\Oa$-improvement 
when one deals with non-zero
quark masses in 
lattice NRQCD. Phrased in the language which is usually
used, the coefficients in the 
NRQCD Lagrangian (or fields) have to be taken as functions
of the NRQCD mass, the bare coupling {\em and $a\mq$}.
Since the dependence on $a\mq$ is needed only in the context of
$\Oa$-improvement, the NRQCD coefficients may be Taylor-expanded in that
variable. Numerically, such terms might be relevant when one deals 
with the
strange quark, since NRQCD is usually applied at rather large lattice spacings.

%% file: sect3.tex
\section{Static quarks in the \SF  \label{s_SF}}

In this section, we first introduce the \SF in the continuum
following~\cite{SF:LNWW}. As a new element we include
static quarks. In~\sect{s_SF_latt} we then define the \SF on the lattice.

\subsection{The \SF   \label{s_SF_basic}}

The space-time topology of the \SF is a finite cylinder with spatial
size $L\times L\times L$ and
with time extent $T$. In the three space directions, periodic boundary
conditions are used for the gauge field, whereas the quark field
boundary conditions are given by \eq{e_lambdabound}. Dirichlet boundary
conditions are chosen at the $x_0=0$ and $x_0=T$ boundaries. In the
following, all space integrals run from $0$ to $L$, all time integrals
run from $0$ to $T$.

For the gauge field $A_k(x)$, boundary fields $C(\vecx)$ and
$C'(\vecx)$ are introduced, such that
\be
  A_k(x)|_{x_0=0}=C(\vecx),\qquad
  A_k(x)|_{x_0=T}=[\projector C'](\vecx)\,,
\ee
where $\projector$ projects onto the gauge invariant content of $C'(\vecx)$
(see \cite{SF:LNWW}). 
No boundary conditions are imposed on the time component of the gauge
field.

Concerning the light quark field, only half of its components are
fixed at the boundaries,
\be
  P_{+}\light(x)|_{x_0=0}=\rhol(\vecx),\quad
  P_{-}\light(x)|_{x_0=T}=\rholprime(\vecx),
\ee
and
\be
  \lightb(x)P_{-}|_{x_0=0}=\rholb(\vecx),\quad
  \lightb(x)P_{+}|_{x_0=T}=\rholbprime(\vecx).
\ee
For consistency, the boundary functions must be such that the
complementary components $P_{-}\rhol,\ldots,\rholbprime P_{-}$ vanish.

The fermionic action is~\cite{SF:stefan1}
\bes
  \label{e_LightSFCont}
  \Sf[A,\lightb,\light] & = & \int\rmd x_0\,\int\rmd^3\vecx\,
    \lightb(x)(\gamma_\mu D_\mu+m)\light(x) \nonumber\\
    & & -\int\rmd^3\vecx\,[\lightb(x)P_{-}\light(x)]_{x_0=0} \nonumber\\
    & & -\int\rmd^3\vecx\,[\lightb(x)P_{+}\light(x)]_{x_0=T}.
\ees

\subsection{Including  static quarks  \label{s_SF_stat}}

As in the case of relativistic quarks, boundary fields $\rhoh$ and
$\rhohbprime$ are defined such that
\be
  \heavy(x)|_{x_0=0}=\rhoh(\vecx),\quad
  \heavyb(x)|_{x_0=T}=\rhohbprime(\vecx).
\ee
Note that no projectors are necessary here because
of \eq{e_constraint}. For the same reason, we have
$P_{-}\heavy(x)=0$. Spatial boundary conditions do not need to be discussed
since the static quarks do not propagate in space. 

Corresponding to~(\ref{e_LightSFCont}), a boundary term has to be
added to the action:
\bes  
  \label{e_HeavySFCont} 
  S_{\rm h}[A,\heavyb,\heavy] = \int\rmd x_0\,\int\rmd^3\vecx\,
    \heavyb(x)D_0\heavy(x)
    -\int\rmd^3\vecx\,[\heavyb(x)\heavy(x)]_{x_0=T}. \nonumber \\
\ees
The complete action is then
\be
  \label{e_SFActCont}
  S[A,\lightb,\light,\heavyb,\heavy]=S_{\rm G}[A]
  +\Sf[A,\lightb,\light]+S_{\rm h}[A,\heavyb,\heavy],
\ee
and the \SF is defined as the partition function
\bes
  \label{e_SFDefCont}
  & & \mathcal{Z}[C',\rholbprime,\rholprime,\rhohbprime;
    C,\rholb,\rhol,\rhoh]= \nonumber \\
  & & \qquad\int D[A]\,D[\light]\,D[\lightb]\,
    D[\heavy]\,D[\heavyb]\,e^{-S[A,\lightb,\light,\heavyb,\heavy]},
\ees
which is a functional of the boundary fields.
In order to have simple renormalization properties, expectation values
of operators are defined for vanishing boundary gauge fields~\cite{impr:pap1}:
\bes
  \label{e_ExpValSF}
  \langle\mathcal{O}\rangle & = & \{{1\over\mathcal{Z}}
    \int D[A]\,D[\light]\,D[\lightb]\,D[\heavy]\,D[\heavyb] \\
    & & \qquad\mathcal{O}e^{-S[A,\lightb,\light,\heavyb,\heavy]}\}|_{
    \rholbprime=\rholprime=\rhohbprime=\rholb=\rhol=\rhoh=0}.
\ees
Here, $\mathcal{O}$ can --- apart from the gauge field and the quark and
anti-quark fields --- contain the light quark ``boundary fields''
\bes
  \label{e_ZetaLightDef}
  \zetal(\vecx)={\delta\over{\delta\rholb(\vecx)}}, & \qquad
  & \zetalb(\vecx)=-{\delta\over{\delta\rhol(\vecx)}}, \\
  \zetalprime(\vecx)={\delta\over{\delta\rholbprime(\vecx)}}, & \qquad
  & \zetalbprime(\vecx)=-{\delta\over{\delta\rholprime(\vecx)}},
\ees
and the corresponding heavy ones
\bes
  \label{e_ZetaHeavyDef}
  \zetahprime(\vecx)={\delta\over{\delta\rhohbprime(\vecx)}}, & \qquad
  & \zetahb(\vecx)=-{\delta\over{\delta\rhoh(\vecx)}}.
\ees

\subsection{Some observables   \label{s_SF_obs}}

For later use, we define the correlation functions
\be
  \fastat(x_0)=-{1\over 2}\int\rmd^3\vecy\,\rmd^3\vecz\,
  \langle\Astat(x)\zetahb(\vecy)\gamma_5\zetal(\vecz)\rangle
\ee
and
\be
  \fonestat=-{1\over{2L^6}}\int
  \rmd^3\vecu\,\rmd^3\vecv\,\rmd^3\vecy\,\rmd^3\vecz\,
  \langle\zetalbprime(\vecu)\gamma_5\zetahprime(\vecv)
  \zetahb(\vecy)\gamma_5\zetal(\vecz)\rangle\,,
\ee
and in particular the ratio
\be
  X(g_0,L/a)={{\fastat(T/2)}\over{\sqrt{\fonestat}}},
\ee
which also depends on $T/L$ and $\theta$.

{\em Renormalization}.\\
We assume that the \SF is finite after adding local counterterms of
dimension $d\leq 4$ to the bulk action, and surface terms of dimension
$d\leq 3$. 
The first part was discussed already in \sect{s_lattice}. 
In complete analogy to the discussion
in \cite{SF:stefan2}, the surface terms can be shown to
be equivalent to a multiplicative renormalization of the
fermionic
boundary fields $\zetal,\,\ldots,\,\zetahprime$, viz.
\be
  {\zetal}_{\rm R}=Z_{\rm l}\zetal,\ldots,
  {\zetahprime}_{\rm R}=Z_{\rm h}\zetahprime\,\,.
\ee
The mass counterterm ($\dmstat$) leads
to an overall exponential factor in the renormalized correlation
functions $\fastatren$ and $\fonestatren$, which cancels in the ratio
$X$. In this ratio, also the wave function renormalization constants
cancel, such that a renormalized ratio $\Xren$ can be written as
\be
  \Xren=\zastat X.
\ee

\subsection{Lattice regularization   \label{s_SF_latt}}

The discretization of the \SF with nonzero boundary gauge fields is
described in~\cite{SF:LNWW,impr:pap1}. Here we choose $C=C'=0$,
leading to the boundary conditions
\be
  U(x,k)|_{x_0=0}=1\,,\quad
  U(x,k)|_{x_0=T}=1
\ee
for the lattice gauge field. The time components remain unconstrained.

For the quark fields, boundary fields $\rhol,\rholb,\rhoh$ and
$\rholprime,\rholbprime,\rhohbprime$ with the same projection
properties as in the continuum are defined on the lattice sites with
$x_0=0$ and $x_0=T$, respectively. The gauge field part of the action,
as well as the light quark action, is given in~\app{a_conv}.

Defining
\be
  \heavy(x)=0\quad\hbox{if $x_0<0$ or $x_0\geq T$},
\ee
the lattice action for the heavy quark with \SF boundary conditions
can be written as
\be
  \label{e_HeavySFLatt}
  S_{\rm h}[U,\heavyb,\heavy] = a^4\sum_x
    \heavyb(x)\nabstar{0}\heavy(x).
\ee

With boundary derivatives defined as in~(\ref{e_ZetaLightDef})
and~(\ref{e_ZetaHeavyDef}), we can write down a discretized version of
our correlation functions,
\be
  \fastat(x_0)=-a^6\sum_{\vecy,\vecz}{1\over 2}
  \langle\Astat(x)\zetahb(\vecy)\gamma_5\zetal(\vecz)\rangle
\ee
and
\be
  \fonestat=-{{a^{12}}\over{L^6}}\sum_{\vecu,\vecv,\vecy,\vecz}{1\over 2}
  \langle\zetalbprime(\vecu)\gamma_5\zetahprime(\vecv)
  \zetahb(\vecy)\gamma_5\zetal(\vecz)\rangle.
\ee

{\em Symanzik improvement}.\\
To cancel $\Oa$ effects, local operators of dimension $4$, which are
summed over the boundary lattice points at $x_0=0$ and $x_0=T$, have to
be added both to the light quark action~\cite{SF:LNWW,impr:pap1} and,
in principle, to the heavy quark action. Writing down all such operators
that are invariant under discrete 3-dimensional Euclidean rotations,
it becomes clear that they can be disregarded in the static case,
as they are zero either because of
\eq{e_constraint} or by formal application of the field equations
\eq{e_eom}.

To achieve complete $O(a)$ improvement, the axial current improvement term
$\castat\fdeltaastat$ with
\be
  \fdeltaastat(x_0)=-a^6\sum_{\vecy,\vecz}{1\over 2}
  \langle\lightb(x)\gamma_j\gamma_5
  \frac12(\lnab{j}+\lnabstar{j})\heavy(x)
  \zetahb(\vecy)\gamma_5\zetal(\vecz)\rangle
\ee
has to be added to $\fastat$ (see section~\ref{s_lattice}).
We therefore define the (improved) ratio,
\be
  \label{e_XIdef}
  X_{\rm I}(g_0,L/a)={{\fastat(T/2)+a\castat\fdeltaastat(T/2)}
    \over{\sqrt{\fonestat}}}\,.
\ee
Note again that the wave function renormalization constants and the
heavy quark mass counterterm cancel in this ratio. Apart from its
arguments, $X_{\rm I}$ also depends on the ratio $T/L$ and the angle
$\theta$.

\subsection{Perturbative expansion   \label{s_PT}}

First of all, we introduce a gauge vector field $q_\mu (x)$, which is
the fluctuation around the background gauge field (zero in our case),
and write the link variables as
\be
  U(x,\mu)=\exp\{g_0aq_\mu(x)\}.
\ee
We then choose the Feynman gauge by the gauge fixing procedure
described in~\cite{impr:pap2}, and add a ghost field term to the action,
which however does not contribute to our correlation functions at one
loop level.

To achieve a perturbative expansion of $X$, the correlation
functions $\fastat$ and $\fonestat$ are expanded in powers of the bare
coupling,
\be
  \fastat(x_0)={\fastat}^{(0)}(x_0)+g_0^2{\fastat}^{(1)}(x_0)+O(g_0^4),
\ee
and
\be
  \fonestat={\fonestat}^{(0)}+g_0^2{\fonestat}^{(1)}+O(g_0^4).
\ee
We use the strategy of~\cite{impr:pap2} to compute the expansion
coefficients. First the Grassmann integrals over the fermionic
variables are carried out, and then the resulting gauge field
dependent observables are expanded in powers of $g_0$. The result of
the first step is conveniently written in terms of the matrices
\be
  \Hlight(x)=a^3\sum_{\vecy}{{\delta\psi_{\rm l,cl}(x)}
    \over{\delta\rhol(\vecy)}}
\ee
and
\be
  \Klight=\cttilde{{a^3}\over{L^3}}\sum_{\vecx}
    \{P_{+}U(x,0)^{-1}\Hlight(x)\}_{x_0=T-a},
\ee
where $\psi_{\rm l,cl}$ is the classical solution of the
Dirac equation, and $\cttilde$ is the coefficient of the boundary
$O(a)$ improvement term in the action~\cite{impr:pap1}.
Corresponding matrices are defined for the static quark,
\be
  \Hheavy(x)=a^3\sum_{\vecy}{{\delta\psi_{\rm h,cl}(x)}
    \over{\delta\rhoh(\vecy)}},
\ee
and
\be
  \Kheavy={{a^3}\over{L^3}}\sum_{\vecx}
    \{P_{+}U(x,0)^{-1}\Hheavy(x)\}_{x_0=T-a}.
\ee
Then the correlation functions can be written as
\be
  \label{e_fastatExpand}
  \fastat(x_0)=-{1\over 2}\langle\tr\{\Hlight(x)^{\dagger}\gamma_0
    \Hheavy(x)\}\rangle_G
\ee
and
\be
  \label{e_fonestatExpand}
  \fonestat={1\over 2}\langle\tr\{\Klight^{\dagger}\Kheavy\}\rangle_G.
\ee
In these formulae, the trace is taken over the Dirac and colour
indices, and the symbol $\langle\ldots\rangle_G$ means the expectation
value with the effective gauge field measure (including the fermionic
determinant).

The matrices $\Hlight$ and $\Klight$ are expanded as
in~\cite{impr:pap2,impr:pap5}. The heavy quark matrices are written as
\be
  \Hheavy(x)=\Hheavy^{(0)}(x)+g_0\Hheavy^{(1)}(x)
    +g_0^2\Hheavy^{(2)}(x)+O(g_0^3)
\ee
and
\be
  \Kheavy=\Kheavy^{(0)}+g_0\Kheavy^{(1)}
    +g_0^2\Kheavy^{(2)}+O(g_0^3)
\ee
with
\be
  \Kheavy^{(0)}={{a^3}\over{L^3}}\sum_{\vecx}P_{+}
    \Hheavy^{(0)}(T-a,\vecx),
\ee
\be
  \Kheavy^{(1)}={{a^3}\over{L^3}}\sum_{\vecx}P_{+}\{\Hheavy^{(1)}(x)
    -aq_0(x)\Hheavy^{(0)}(x)\}_{x_0=T-a},
\ee
\be
  \Kheavy^{(2)}={{a^3}\over{L^3}}\sum_{\vecx}P_{+}\{\Hheavy^{(2)}(x)
    -aq_0(x)\Hheavy^{(1)}(x)+\frac12[aq_0(x)]^2\Hheavy^{(0)}(x)\}_{x_0=T-a}\,.
  \nonumber \\
\ee
The exact form of $\Hheavy^{(0)}$, $\Hheavy^{(1)}$, and
$\Hheavy^{(2)}$
is given in \App{a_StaticFeyn}. Inserting these expansions up to $O(g_0^2)$
into~(\ref{e_fastatExpand}) and~(\ref{e_fonestatExpand}), we get the
expansion of $\fastat$ and $\fonestat$ in terms of one-loop diagrams,
which can be found in \App{a_StaticFeyn}.

Using furthermore
\be
 \fdeltaastat= g_0^2{\castat}^{(1)}{\fdeltaastat}^{(0)}+O(g_0^4),
\ee
with ${\castat}^{(1)}$ taken from~(\ref{castat}),
the expansion coefficients of $X_{\rm I}$,
\eq{e_XIdef}, defined by
\be
  X_{\rm I}=X_{\rm I}^{(0)}+g_0^2X_{\rm I}^{(1)}+O(g_0^4)
\ee
can be calculated. Their exact form is
\be
  X_{\rm I}^{(0)}(L/a)={{{\fastat}^{(0)}(T/2)}
    \over{\sqrt{{\fonestat}^{(0)}}}},
\ee
\bes
  X_{\rm I}^{(1)}(L/a) & = & 
   {\fastat}^{(1)}(T/2)/\sqrt{{\fonestat}^{(0)}}
   -\frac12 {\fastat}^{(0)}(T/2)\, ({\fonestat}^{(0)})^{-3/2}\,
    {\fonestat}^{(1)} \nonumber\\
  & & +\,a\,{\castat}^{(1)}{\fdeltaastat}^{(0)}(T/2)/\sqrt{{\fonestat}^{(0)}}\,.
\ees

%% file: sect4.tex
\section{The renormalized current in the SF-scheme  \label{s_renorm}}

We are now prepared to discuss the renormalization of the current in 
different schemes. In order to separate clearly the dependence on
the regularization and the scheme dependence due to 
different normalization conditions, we first renormalize the
current by minimal subtraction on the lattice. The resulting current
has finite matrix elements and can be related to the 
one in $\MSbar$-normalization by a finite renormalization.
We then introduce the SF-scheme, which in contrast to
the other two schemes is defined beyond perturbation theory.

\subsection{Lattice minimal subtraction and $\MSbar$-normalization
            \label{s_MSbar}}

The ratio $X$ can be
made finite by a multiplicative renormalization of the coupling, the
light quark mass, and the static-light axial current,
\bes
  g_{\rm R}&=&Z_{\rm g}(1+b_{\rm g}am_{\rm q})g_0, 
  \\
  m_{\rm R}&=&Z_{\rm m}(1+b_{\rm m}am_{\rm q})m_{\rm q},
  \\
  X_{\rm R}&=&\zastat(1+\bastat am_{\rm q})X.
\ees
The parameter $m_{\rm q}$ denotes the subtracted mass of the light
quark,
\be
  m_{\rm q}=m_0-m_{\rm c},
\ee
where $m_0$ is the bare light quark mass. At finite lattice spacing,
different definitions of the critical mass $m_{\rm c}$ are possible,
differing by
${\rm O}(a^2)$ terms.
Our choice is to introduce the PCAC mass as defined
in~\cite{impr:pap1}, and then one can calculate the bare quark mass at which 
the 
PCAC mass vanishes.
At each order of perturbation theory,
this mass can be extrapolated to the continuum limit, and it is this
limit that we will use as the critical light quark mass in our calculations.
Note that $X$ is only multiplied by
the axial current renormalization constant, as the wave function
renormalization constants and the mass dependent wave function
improvement factors cancel.

In perturbation theory, coupling, mass and static-light axial current are
logarithmically divergent when taking the continuum limit. We define
the {\em lattice minimal subtraction scheme} by requiring that the
coefficients of the renormalization constants are polynomials in
$\ln(a\mu)$ (without constant parts) at each order of perturbation theory, 
with some
renormalization scale $\mu$.

We chose $m_0=m_{\rm c}$, which means that we do not have to take care
of the mass renormalization and the mass dependent improvement
terms. Expanding the correlation functions up to $g_0^2$, only the
tree level value $Z_{\rm g}=1$ is needed, and we remain with the
current renormalization constant, which is
\be
  Z_{\rm A,lat}^{\rm stat}=1-\gamma_0\ln(a\mu)\glat^2
    +O(\glat^4),
\ee
with the scheme independent one loop anomalous dimension 
\cite{Shifman:1987sm,Politzer:1988wp}
\be
  \gamma_0=-{1\over{4\pi^2}}.
\ee
The minimally subtracted ratio,
\be
  X_{\rm I,lat}=Z_{\rm A,lat}^{\rm stat}X_{\rm I}\,,
\ee
is finite, and its coefficients,
\be
  \label{e_Xlatexp}
  X_{\rm I,lat}^{(0)}=X_{\rm I}^{(0)}\,,\quad
  X_{\rm I,lat}^{(1)}=X_{\rm I}^{(1)}-\gamma_0\ln(a\mu)X_{\rm I}^{(0)}
\ee
can be extrapolated to the continuum limit $a/L\rightarrow 0$.
Setting $\mu=1/L$, and using the extrapolation procedure 
described in~\cite{Bode:1999sm} we obtained the 
values listed in \tab{t_pert_coeff}.

\input{tab_pert_coeff}

The lattice minimal subtraction scheme is related to other 
schemes by finite
renormalization. The most widely used scheme is the 
$\msbar$ scheme of dimensional regularization. In the present context
this means that the renormalized coupling is defined 
by the $\msbar$ prescription, but the renormalization of
the current in the effective theory is fixed by matching to
the current in the full theory (the normalization of
the latter is fixed uniquely by current algebra).
``Matching'' means that matrix elements of the current in the 
effective theory are equal to those in the full theory up
to terms suppressed by inverse powers of the heavy quark mass.
For more details we refer the reader to \cite{Ji:1991pr,Neubert:1994mb}.
We denote the current normalized in this way by
$\ArenMSbar$. Its relation to the lattice minimally 
subtracted current,
\be
 \ArenLat = Z_{\rm A,lat}^{\rm stat} \Astat\,,
\ee
is
\bes
  \label{e_MSbarLat}
  \ArenMSbar&=&\chiAstatMSbarLat\ArenLat\,,\\  
   \chiAstatMSbarLat&=&1
    +{\chiAstatMSbarLatOne}\gbarLat^2+O(\gbarLat^4). \nonumber
\ees
Here and below, all scale-dependent renormalized quantities are
taken at the same renormalization scale $\mu$.
Borrelli and Pittori have computed $\chiAstatMSbarLatOne$
by considering matrix elements between quark states, and
regulating infrared divergences by a gluon mass \cite{BorrPitt}. Extracting 
from their results the terms which contribute to the
(local) current, \eq{e_StatAxial}, one finds
\be
  \label{e_zaBo}
  {\chiAstatMSbarLatOne}={1\over 6\pi^2}-0.14.
\ee
In a separate publication we will report on a computation
of the matching between effective theory and full theory,
in lattice regularization and for infrared 
finite correlation functions (defined in the \SF) \cite{xy:zastat}. 
Here we already quote a result which 
is directly relevant in the present context,
\be
  \label{e_zaMSbar}
  {\chiAstatMSbarLatOne}={1\over 6\pi^2}-0.137(1)\,.
\ee
It is in agreement with \eq{e_zaBo}.
The lattice minimal subtraction scheme
does of course depend on the details of the chosen discretization.
\Eq{e_zaMSbar} holds if the
\emph{improved} Wilson action is used for the light quark. 
Results for 
the unimproved case can be found 
in~\cite{stat:boucaud_za,stat:eichhill_za}.

\subsection{The SF-scheme\label{s_scheme}}

The SF scheme is a renormalization scheme using observables defined
with Schr\"odinger Functional
boundary conditions, and identifying the renormalization scale with the
inverse box size $L^{-1}$.
First of all, a renormalized coupling
$$
  \gbarSF^2(L)
$$
is introduced, the definition of which can be found in~\cite{alpha:SU3}
and the masses of the light quarks are set to zero.
We then define the renormalized
static axial current in the SF scheme,
\be
 \ArenSF(\mu) = \zastatSF\,(1+\bastat a\mq) \Astatimpr
\ee
by using the ratio \eq{e_XIdef}, imposing the renormalization 
condition
\be
  \label{e_SFdef}
  \zastatSF(g_0,L/a) X_{\rm I}(g_0,L/a)=X^{(0)}(L/a) \;\;{\rm at }\; m_0=\mc
\ee
and setting $\mu=1/L$.

Perturbatively, the finite renormalization constant connecting the
renormalized currents in the SF and $\msbar$ schemes can be calculated,
\be
  \label{e_SFMSbar}
  \ArenSF=\chiAstatSFMSbar\, \ArenMSbar,
\ee
\be
  \chiAstatSFMSbar=1+\gbarMSbar^2{\chiAstatSFMSbar}^{(1)}+O(\gbarMSbar^4).
\ee
Combining eqs. (\ref{e_Xlatexp},\ref{e_MSbarLat},\ref{e_SFdef}) and \eq{e_SFMSbar},
and expanding in the coupling, we get
\be
  {\chiAstatSFMSbarOne}=-{\chiAstatMSbarLatOne}
    -{{X_{\rm I,lat}^{(1)}}\over{X^{(0)}}},
\ee
where it is understood that the last term is to be evaluated for
$a/L\rightarrow 0$. The results are contained in \tab{t_pert_coeff}.

\section{Anomalous dimension in the SF-scheme \label{s_gamma}}

To match perturbative and non-perturbative results with sufficient
precision, one needs the
static-light axial current's anomalous dimension $\gamma(\gbar)$,
defined by the renormalization group equation
\be
  \mu{{\partial\Aren}\over{\partial\mu}}=\gamma(\gbar)\Aren,
\ee
at two loop order in the SF scheme. We calculate the coefficient
$\gamSF_1$ in the expansion
\be
  \gamSF(\gbarSF)=-\gbarSF^2(\gamma_0+\gamSF_1\gbarSF^2+\ldots)
\ee
by conversion from the $\msbar$ scheme to the
SF scheme. 

The two loop anomalous dimension in the $\msbar$ scheme is known
from~\cite{Ji:1991pr,BroadhGrozin,Gimenez:1992bf},
\be
  \gamMSbar_1=-{1\over 576\pi^4}({127\over 2}+28\zeta(2)-5N_{\rm f})\,.
\ee
Here $\zeta$ is the Riemann zeta function, and $N_{\rm f}$ is the
number of relativistic quark flavours.

The renormalized couplings in the two
schemes are related by
\be
  \gbarSF^2=\chi_{\rm g}\gbarMSbar^2, \qquad
  \chi_{\rm g}=1+\chi_{\rm g}^{(1)}\gbarMSbar^2+O(\gbarMSbar^4),
\ee
were the 1-loop coefficient is
\be
  \left.\chi_{\rm g}^{(1)}\right|_{\mu=1/L}=-{1\over 4\pi}(c_{1,0}+c_{1,1}N_{\rm
    f})
\ee
with~\cite{impr:pap2,pert:1loop}
\be
  c_{1,0}=1.25563(4),\qquad c_{1,1}=0.039863
\ee
for the SF-coupling defined
at $\theta=\pi/5$. The relation between the currents in the two schemes
is known from \tab{t_pert_coeff}.

Analogous to~\cite{mbar:pert}, where the discussion is carried through for
the anomalous dimension of quark masses, we have
\be
  \gamSF_1=\gamMSbar_1+2b_0\chi_{\rm A}^{{\rm stat}(1)}-
  \gamma_0\chi_{\rm g}^{(1)},
\ee
where
\be
  b_0={1\over{(4\pi)^2}}(11-{2\over 3}N_{\rm f})
\ee
is the universal one loop coefficient of the $\beta$ function.

From our result for $\chiAstat^{(1)}$ in the $T=L$ case,
\tab{t_pert_coeff}, we get
\bes
  \gamSF_1(\theta=0.0) & = &
  {1\over{(4\pi)^2}}\{0.52(2)-0.0733(13)N_{\rm f}\}, \\
  \gamSF_1(\theta=0.5) & = &
  {1\over{(4\pi)^2}}\{0.08(2)-0.0466(13)N_{\rm f}\}, \\
  \gamSF_1(\theta=1.0) & = &
  {1\over{(4\pi)^2}}\{-0.36(2)-0.0199(13)N_{\rm f}\}\,.
\ees
For the range of $\theta$ considered, these coefficients are rather small
and the perturbative series appears to be well behaved.

\section{Lattice spacing effects in the SF-scheme
         \label{s_ssf}}

The renormalization constants at energy scales $1/L$ and $1/(2L)$ are
connected by the \emph{step scaling function},
\be
  \SigmaAstat(u,a/L)={{\zastatSF(g_0,2L/a)}\over
  {\zastatSF(g_0,L/a)}}\qquad {\rm at}\quad \gbarSF^2(L)=u\,.
\ee
This lattice step scaling function has a continuum limit $\sigmaAstat$,
\be
  \SigmaAstat(u,a/L)=\sigmaAstat(u)+\rmO(a/L).
\ee
With $\sigmaAstat$, a non-perturbative renormalization group
can be constructed without taking reference to the lattice regularization.

\begin{figure}
  \begin{center}
  \epsfig{file=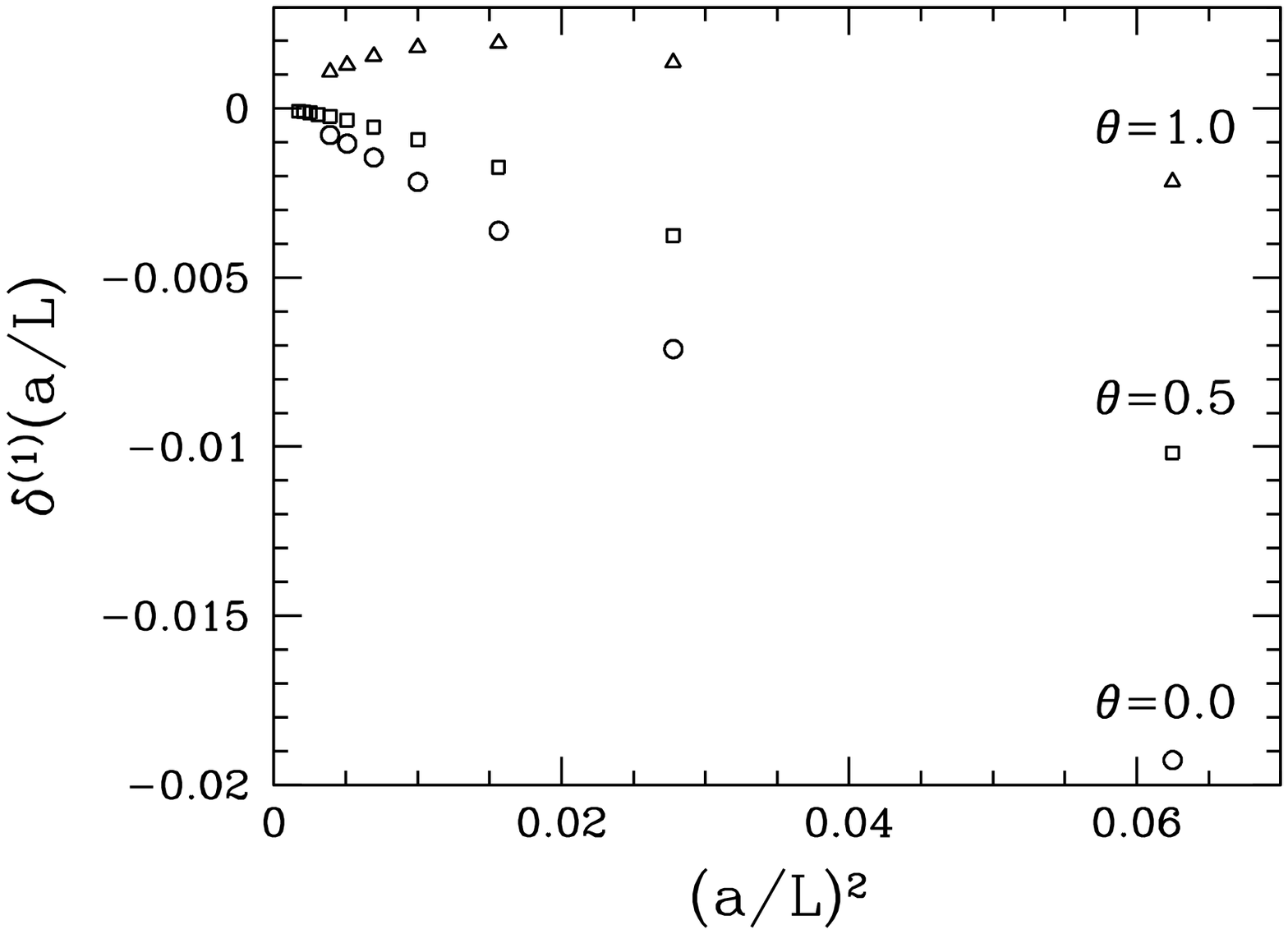,width=9cm,clip=}
  \end{center}
  \caption{Discretization errors in the $\Oa$ improved  
    step scaling function at one loop order of perturbation
    theory. Points are for $L/a=4,6,8,\ldots$.}
  \label{f_deltaplot}
\end{figure}

As $\SigmaAstat(u,a/L)$ will be the central object of
non-perturbative studies, it is interesting to investigate its
discretization errors at the one loop level. As an
estimate for these errors, we have calculated
\bes
  \delta(u,a/L) & \equiv & {\latstep(u,a/L)- \contstep(u) \over 
                               \contstep(u)} 
=0 + \delta^{(1)}(a/L)u + \ldots
\ees
for three values of $\theta$. 
The results are displayed in
\fig{f_deltaplot}, showing that lattice spacing effects -- which are 
$\rmO((a/L)^2)$ since we use $\Oa$ improvement -- are moderate at one loop
order. Again, we have used $\mc$ extrapolated to $a\to0$ in this exercise.
In MC computations one has to use a definition of $\mc$ at finite $a/L$ 
instead. Then lattice spacing errors are somewhat different. We have checked 
that in the case at hand these differences are tiny when typical choices for
$\mc$, such as the one in \cite{mbar:pap1}, are made.

\section{The improved axial current at 1-loop order \label{s_impr}}

Throughout our calculations, we have used the value from \eq{castat} for
the current improvement coefficient at one loop order, obtained by
Morningstar and Shi\-ge\-mitsu~\cite{MORNINGSTAR1} using NRQCD methods.
We have also computed
this coefficient from our data for $\theta=0.5$ and $\theta=1.0$.
First of all, we calculated the ratio
$X$ \emph{without} the current improvement term, and then extracted its 
$\Oa$ part by the fitting procedure described in~\cite{Bode:1999sm},
using the formula given in \App{a_StatImpr}.
The results are
\be
  {\castat}^{(1)}=-{1\over{4\pi}}\times 1.01(7)\qquad {\rm
    at}\quad\theta=0.5
\ee
and
\be
  {\castat}^{(1)}=-{1\over{4\pi}}\times 1.01(5)\qquad {\rm
    at}\quad\theta=1.0,
\ee
in agreement with \eq{castat}. We note that our values also
agree with the more recent and precise calculation 
of~\cite{Ishikawa},
\bes
{\castat}^{(1)}= -{1\over{4\pi}}\times 1.0351(1)\,.
\ees

From a similar analysis of data obtained with
a nonzero light quark mass, the coefficient $\bastat$ introduced in
\sect{s_Oa} can be calculated at one-loop order. Our result, computed
from \eq{e_BastatForm} for $\theta=0$ and $z\equiv Lm_{\rm lat}=1$, is
\be
  \label{bastat1}
  {\bastat}=\frac12 + {\bastat}^{(1)} g_0^2 + \ldots, \,\quad 
  {\bastat}^{(1)}=-0.056(7).
\ee
Here the error is due to the uncertainty in the $a/L\rightarrow 0$
limit. Determinations with other values of $z$ and $\theta$ turned out 
to be entirely consistent with \eq{bastat1}.

%% file: tab_pert_coeff.tex
\begin{table}[htb]
\centering
\begin{tabular}{l l l l }
\hline \\[-2.0ex]
$\theta$ & $X_{\rm I,lat}^{(0)}$ & $X_{\rm I,lat}^{(1)}$ & ${\chiAstatSFMSbar}^{(1)}$
\\[1.0ex]
\hline \\[-2.0ex]
  0.0 & $-1$.7320508076 & $-0$.110337(3) & 0.056(1) \\
  0.5 & $-1$.6037987399 & $-0$.135646(2) & 0.036(1) \\
  1.0 & $-1$.4193852065 & $-0$.147875(1) & 0.016(1) \\[1.0ex]
\hline\\
\end{tabular}
\caption{\footnotesize Perturbative coefficients for $T=L$ and various $\theta$ 
in the continuum limit $a/L=0$.
\label{t_pert_coeff}}
\end{table}

%% file: sect5.tex
\section{Summary   \label{s_concl}}

Our paper covers two main topics.
The first one concerns the formulation of the static quark effective theory
on the lattice, in particular the question of $\Oa$ improvement. 
The second topic is of a more practical nature and deals with the 
formulation of a renormalization condition for the static light axial
current which can be used to renormalize the current non-perturbatively 
by MC computations. 

       I. We have started from the general assumption that the divergence structure
       of static quark effective field theory 
       can be inferred from simple dimensional counting and that the 
       theory can be made finite by addition of local counterterms.

         Clearly it would desirable to have a proof that this
             assumption is correct to all orders of perturbation theory.
             In absence of a proof, we may, however, perform checks through
             perturbative computations as well as finally MC computations.
             All results of our one loop computations are in agreement with
             the assumption. Particularly convincing is the agreement
             of the renormalization factor \eq{e_zaBo} of
             \cite{BorrPitt} with our result \eq{e_zaMSbar}, obtained
             from a completely different quantity. A similar
             statement holds for the 
             renormalization beyond the leading order, i.e. $\Oa$-improvement:
             the coefficient $c^{\rm stat}_{\rm A}$ from
             \cite{MORNINGSTAR1,Ishikawa} yields $\Oa$-improvement for our 
             correlation 
             functions.

       In more detail, the following terms are found to be necessary and sufficient for
       renormalization and improvement of the theory. 
   \begin{itemize}\itemsep 1pt
      \item[$\circ$] In the Lagrangian only a mass counterterm $\dmstat \heavyb(x) \heavy(x)$
            is necessary at the leading order in the lattice spacing (i.e. to
            obtain a finite continuum limit).
      \item[$\circ$] The static light axial current is renormalized multiplicatively. 
      \item[$\circ$] For vanishing light quark masses, the static sector defined
            with the action of Eichten and Hill, \eq{e_LatLag}, is $\Oa$-improved.
            At finite quark mass, a correction term 
            $\delta S_{\rm h} = a^5 \sum_x \,\bstat \, \mq^2 \, \heavyb(x)  \heavy(x)$
            has to be added to achieve $\Oa$-improvement. Just like $\dmstat$, such a 
            term affects 
            correlation functions only in a trivial manner.
      \item[$\circ$] Improvement of axial and vector current
necessitates two
\linebreak improvement
            terms each. For example the renormalized axial current may be written
            as 
        $$\Aren = \zastat\,(1+\bastat a\mq)\left[\Astat +  a\castat \lightb\gamma_j\gamma_5
       \frac12(\lnab{j}+\lnabstar{j})\heavy\right]\,,$$ with $\zastat$ the renormalization
         constant defined at vanishing quark mass and two improvement constants 
         $\castat,\bastat$. In perturbation theory, we have
         \bes
             \bastat &=& \frac12 + {\bastat}^{(1)} +\rmO(g_0^4) \nonumber \\
             \castat &=& {c^{\rm stat}_{\rm A}}^{(1)}g_0^2+{\rm O}(g_0^4)\,.
         \ees
         Indeed,  ${c^{\rm stat}_{\rm A}}^{(1)}$ had already been determined 
         in the literature \cite{MORNINGSTAR1,Ishikawa} and was confirmed
         once again by us. It is rather small. The one loop coefficient 
         ${\bastat}^{(1)}$ is given in \eq{bastat1}. 
       \item[$\circ$] Conveniently, the renormalization of the \SF with static quarks
         needs only a multiplicative renormalization of the ``boundary
\linebreak fields''. No
         new $\Oa$ (surface-) counterterms appear when the light quark mass
         is set to zero.
   \end{itemize}
       Apart from the special situation of the \SF, the renormalization 
       properties of the effective theory have been known 
       already and are 
       included above only for completeness. On the other hand, the 
       general structure of $\Oa$-improvement has not been derived before. 
       We remark that
       terms such as $\bastat$ are necessary also in NRQCD, see end 
       of \sect{s_lattice}.

   II. We have introduced a renormalization condition for $\zastat$ in the 
       framework of the \SF. This is done such that a non-perturbative computation
       of the scale dependence as well as the value of $\zastat$ at a particular
       value of the renormalization scale are possible, in complete analogy 
       to the case of the QCD quark masses \cite{mbar:pap1}.
       \begin{itemize}\itemsep 1pt
       \item[$\circ$] In order to be able to follow the strategy of \cite{mbar:pap1}, 
         the two loop anomalous
       dimension has been computed; it is given at the end of \sect{s_renorm}.
       \item[$\circ$] Lattice spacing effects in the step scaling function 
       (describing the scale 
       dependence of the current) have been computed to one loop. They are
       small at this order, promising the possibility of controlled continuum
       extrapolations of future MC results.
   \end{itemize}
MC computations of the step scaling function are in progress \cite{zastat:nonpert}. 

As a future perspective, 
one can follow the same kind of strategy also to 
renormalize the  operators responsible for B$\bar{\rm B}$ mixing in the
static approximation. While this is very attractive in particular in view of
the phenomenological relevance, one should be aware that new 
difficulties will appear: in this case there is
a scale-dependent mixing 
and more terms will be necessary to achieve $\Oa$-improvement. 

{\bf Acknowledgement.} We would like to thank Peter Weisz, Ulli Wolff and
   Stefan Sint for
useful discussions. Particular thanks go to Martin L\"uscher who's  FORTRAN programs 
\cite{impr:pap2} were the basis for our programs computing
the one-loop correlation functions.

%% file: appa.tex
\section{Notations and action for relativistic quarks and gauge fields
         \label{a_conv}}
We use the conventions of \cite{impr:pap1}. Together with a few
extensions they are listed here 
for the readers convenience.

\subsection{Index conventions}

Lorentz indices $\mu,\nu,\ldots$ are taken from the middle of the 
Greek alphabet and run from 0 to 3. Latin indices $k,l,\ldots$
run from 1 to 3 and 
are used to label the components of spatial vectors.  
For the Dirac indices capital letters $A,B,\ldots$ from the 
beginning of the alphabet are taken. They run from 1 to 4.
Colour vectors in the fundamental representation of SU($N$)
carry indices $\alpha,\beta,\ldots$ ranging from 1 to $N$, 
while for vectors in the adjoint representation, Latin indices
$a,b,\ldots$ running from 1 to $N^2-1$ are employed.

Repeated indices are always summed over unless otherwise
stated and scalar products are taken with euclidean metric.

\subsection{Dirac matrices}

We choose a chiral representation for 
the Dirac matrices, where
\be
  \dirac{\mu}=\pmatrix{0                 & e_{\mu}  \cr
                       e_{\mu}^{\dagger} & 0        \cr}. 
\ee
The $2\times2$ matrices $e_{\mu}$ are taken to be 
\be
  e_0=-1,\qquad e_k=-i\sigma_k,
\ee
with $\sigma_k$ the Pauli matrices.
It is then easy to check that 
\be
  \diracstar{\mu}{\dagger}=\dirac{\mu},\qquad
  \{\dirac{\mu},\dirac{\nu}\}=2\delta_{\mu\nu}.
\ee
Furthermore, if we define
$\dirac{5}=\dirac{0}\dirac{1}\dirac{2}\dirac{3}$, we have
\be
  \dirac{5}=
  \pmatrix{1&0\cr 0&-1\cr}.
\ee
In particular, $\dirac{5}=\diracstar{5}{\dagger}$
and $\diracstar{5}{2}=1$.
The hermitean matrices
\be
  \sigma_{\mu\nu}={i\over2}\left[\dirac{\mu},\dirac{\nu}\right]
\ee
are explicitly given by
\be
  \sigma_{0k}=\pmatrix{\sigma_k&0\cr 0&-\sigma_k\cr},
  \qquad
  \sigma_{ij}=-\epsilon_{ijk}\pmatrix{\sigma_k&0\cr 0&\sigma_k\cr},
\ee  
where $\epsilon_{ijk}$ is the totally anti-symmetric tensor with
$\epsilon_{123}=1$.

\subsection{Lattice conventions}

Ordinary forward and backward lattice derivatives 
act on colour singlet functions 
$f(x)$ and are defined through
\bes
  \drv{\mu}f(x)&=&{1\over a}\bigl[f(x+a\hat{\mu})-f(x)\bigr], \nonumber
  \\
  \drvstar{\mu}f(x)&=&{1\over a}\bigl[f(x)-f(x-a\hat{\mu})\bigr],
  \label{e_deriv}
\ees  
where $\hat{\mu}$ denotes the unit vector in direction $\mu$.
The gauge covariant derivative operators,
acting on a quark field $\psi(x)$, are given by
\bes
  \nab{\mu}\psi(x)&=&
  {1\over a}\bigl[\lambda_{\mu}U(x,\mu)\psi(x+a\hat{\mu})-\psi(x)\bigr],
  \\
  \nabstar{\mu}\psi(x)&=&
  {1\over a}\bigl[\psi(x)-\lambda_{\mu}^{-1}U(x-a\hat{\mu},\mu)^{-1}
  \psi(x-a\hat{\mu})\bigr].
\ees
The phase factors
\be
 \label{e_lambdamu}
  \lambda_{\mu}=\rme^{ia\theta_{\mu}/L},\qquad
  \theta_{0}=0,\quad -\pi<\theta_k\leq\pi,
\ee
are equivalent to imposing the generalized
periodic boundary conditions
\be
  \label{e_lambdabound}
  \psi(x+L\hat{k})=\rme^{i\theta_k}\psi(x),
  \qquad
  \psibar(x+L\hat{k})=\psibar(x)\rme^{-i\theta_k},
\ee
They depend on the spatial 
extent $L$ of the lattice and are all equal to $1$ on 
the infinite lattice.
The left action of the lattice derivative operators
is defined by
\bes
  \psibar(x)\lvec{\nab{\mu}}&=&
  {1\over a}\left[\,
  \psibar(x+a\hat{\mu})U(x,\mu)^{-1}
  \lambda_{\mu}^{-1}-\psibar(x)\,\right],
  \\
  \psibar(x)\lvec{\nabstar{\mu}}&=&
  {1\over a}\left[\,
  \psibar(x)-\psibar(x-a\hat{\mu})U(x-a\hat{\mu},\mu)
  \lambda_{\mu}\,\right].
\ees  
Our lattice version of $\delta$-functions are 
\be
 \label{e_delta}
 \delta(x_\mu) = a^{-1} \delta_{x_\mu 0}\,,\quad
 \delta(\vecx) = \prod_{k=1}^3 \delta(x_k) \,,\quad
 \delta(x) = \prod_{\mu=0}^3 \delta(x_\mu)  
\ee
and we use
\bes
 \label{e_theta}
  \theta(x_\mu) &=& 1 \hbox{ for } x_\mu\geq0 \\
  \theta(x_\mu) &=& 0 \hbox{ otherwise }      \nonumber
\ees

\subsection{Continuum gauge fields}

An SU($N$) gauge potential in the continuum theory 
is a vector field $A_{\mu}(x)$
with values in the Lie algebra su($N$). It may thus be written as
\be
  A_{\mu}(x)=A_{\mu}^a(x)T^a
\ee
with real components $A_{\mu}^a(x)$ and
\be
  (T^a)^\dagger = - T^a,\quad \tr\{T^aT^b\}=-\frac{1}{2}\delta^{ab}.
\ee 
The associated field tensor,
\be
  F_{\mu\nu}(x)=
  \partial_{\mu}A_{\nu}(x)-\partial_{\nu}A_{\mu}(x)
  +[A_{\mu}(x),A_{\nu}(x)],
\ee
may be decomposed similarly and 
the right and left action of the covariant derivative 
$D_{\mu}$ is defined by
\bes
  D_{\mu}\psi(x)&=&
  (\drv{\mu}+A_{\mu}+i\theta_{\mu}/L)\psi(x),
   \\
  \psibar(x)\lvec{D_{\mu}}&=&
  \psibar(x)(\lvec{\drv{\mu}}-A_{\mu}-i\theta_{\mu}/L).
\ees  
The abelian gauge field $i\theta_{\mu}/L$ appearing here 
corresponds to the phase factors $\lambda_{\mu}$, \eq{e_lambdamu}, 
in the lattice theory.

\subsection{Lattice action}
Let us first assume that the theory is defined on an infinite
lattice. 
A gauge field $U$ on the lattice is an assignment of a matrix 
$U(x,\mu)\in\SUn$ to every lattice point $x$ and direction 
$\mu=0,1,2,3$.
Quark and anti-quark fields, $\psi(x)$ and $\psibar(x)$,
reside on the lattice sites and  
carry Dirac, colour and flavour indices.
The (unimproved) lattice action is of the form
\be
  S[U,\psibar,\psi\,]=\Sg[U]+\Sf[U,\psibar,\psi\,],
\ee
where $\Sg$ denotes 
the usual Wilson plaquette action and $\Sf$ the Wilson quark action.
Explicitly we have
\be
  \Sg[U]={1\over g_0^2}\sum_p\tr\{1-U(p)\}
\ee
with $g_0$ being the bare gauge coupling and 
$U(p)$ the parallel transporter around the plaquette $p$.
The sum runs over all {\it oriented}\/ plaquettes $p$
on the lattice. 
The quark action,
\be
  \label{e_Sf}
  \Sf[U,\psibar,\psi\,]=a^4\sum_{x}\psibar(x)(D+m_0)\psi(x),
\ee
is defined in terms of the 
Wilson-Dirac operator
\be
  \label{e_dirop}
  D=\frac{1}{2}\left\{
  \dirac{\mu}(\nabstar{\mu}+\nab{\mu})-a\nabstar{\mu}\nab{\mu}\right\},
\ee
which involves the gauge covariant lattice derivatives 
$\nab{\mu}$ and $\nabstar{\mu}$, \eq{e_deriv}, and 
the bare quark mass, $m_0$.

The improved action is given by \cite{impr:sw,impr:pap1}
\bes
  \label{e_simpr}
    \Simpr[U,\psibar,\psi]&=&S[U,\psibar,\psi]+\delta S[U,\psibar,\psi],
  \\
  \delta S[U,\psibar,\psi]&=&
  a^5\sum_{x}
  \csw\,
  \psibar(x)\frac{i}{4}\sigma_{\mu\nu}\widehat{F}_{\mu\nu}(x)\psi(x),
\ees
with
\bes
    \widehat{F}_{\mu\nu}(x)&=&{1\over8a^2}\left\{
  Q_{\mu\nu}(x)-Q_{\nu\mu}(x)\right\},
  \\
  Q_{\mu\nu}(x)
  &=&\,U(x,\mu)U(x+a\hat{\mu},\nu)
    U(x+a\hat{\nu},\mu)^{-1} U(x,\nu)^{-1}
  \\
  && +\,U(x,\nu)U(x-a\hat{\mu}+a\hat{\nu},\mu)^{-1}
    U(x-a\hat{\mu},\nu)^{-1}U(x-a\hat{\mu},\mu)
  \nonumber \\ && 
  +\,U(x-a\hat{\mu},\mu)^{-1}U(x-a\hat{\mu}-a\hat{\nu},\nu)^{-1}
  \nonumber \\ && \qquad\qquad\qquad\qquad\qquad\quad
   \times U(x-a\hat{\mu}-a\hat{\nu},\mu)U(x-a\hat{\nu},\nu)
  \nonumber \\ && 
  +\,U(x-a\hat{\nu},\nu)^{-1}U(x-a\hat{\nu},\mu)
    U(x+a\hat{\mu}-a\hat{\nu},\nu)U(x,\mu)^{-1}.\nonumber 
\ees
Turning now to the case of \SF boundary conditions,
the pure gauge action on the lattice is \cite{SF:LNWW}
\be  
  \Sg[U]={1\over g_0^2}\sum_p w(p)\tr\{1-U(p)\},
\ee
with a weight $w(p)$ equal to 1 for all $p$ except for
the spatial plaquettes at $x_0=0$ and
$x_0=T$, where $w(p)= \frac{1}{2}$.

The quark action in the \SF can be written exactly as in \eq{e_Sf} when
the conventions
\be
   \psi(x)=0\quad\hbox{if $x_0<0$ or $x_0>T$}
\ee
and
\be
    P_{-}\psi(x)|_{x_0=0}=
  P_{+}\psi(x)|_{x_0=T}=0
\ee
are adopted.

%% file: appb.tex
\section{Static quarks in perturbation theory
  \label{a_StaticPert}}

\subsection{Heavy quark Feynman rules
  \label{a_StaticFeyn}}

Using the lattice action~(\ref{e_HeavySFLatt}) for the heavy quark,
we get the heavy quark propagator
\be
  S_{\rm h}(x,y)=U(x-a\hat{0},0)^{-1}\ldots
  U(y,0)^{-1}\theta(x_0-y_0)\delta(\vecx-\vecy)P_{+},
\ee
with $\theta(x_0)$ and $\delta(\vecx)$ defined in
(\ref{e_theta},\ref{e_delta}).
Its fourier transform $\tilde{S}_{\rm h}$ is defined by
\be
  S_{\rm h}(x,y)={1\over{L^3}}\sum_{\vectp}e^{i\vectp(\vecx-\vecy)}
                      \tilde{S}_{\rm h}(x_0,y_0;\vectp).
\ee
Expanding in terms of $g_0$, we find
\be
  \tilde{S}_{\rm h}^{(0)}(x_0,y_0;\vectp)=P_{+}.
\ee

\begin{figure}
  \noindent
  \begin{center}
  \begin{minipage}[b]{.3\linewidth}
    \centering\epsfig{figure=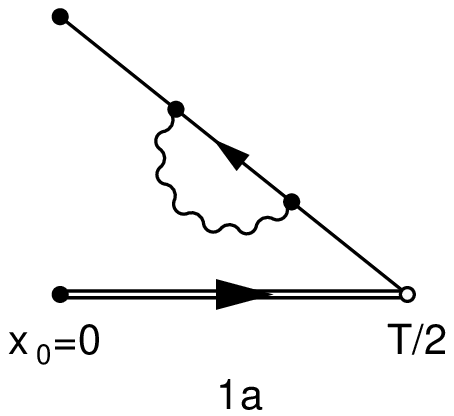,width=\linewidth}
  \end{minipage}
  \begin{minipage}[b]{.3\linewidth}
    \centering\epsfig{figure=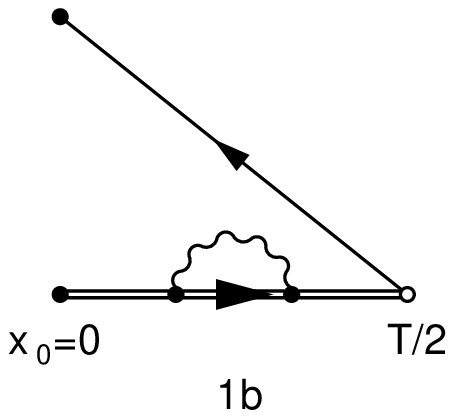,width=\linewidth}
  \end{minipage}\\
  \vspace{4mm}
  \begin{minipage}[b]{.3\linewidth}
    \centering\epsfig{figure=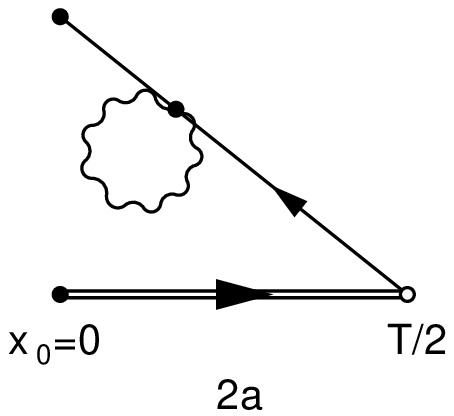,width=\linewidth}
  \end{minipage}
  \begin{minipage}[b]{.3\linewidth}
    \centering\epsfig{figure=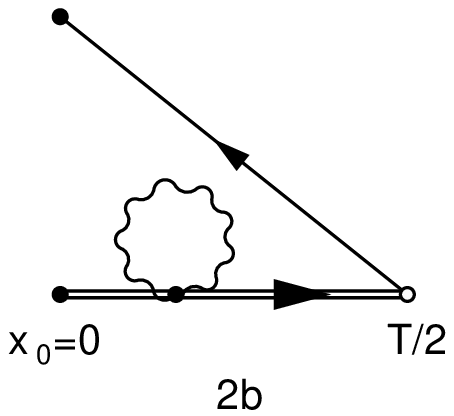,width=\linewidth}
  \end{minipage}
  \begin{minipage}[b]{.3\linewidth}
    \centering\epsfig{figure=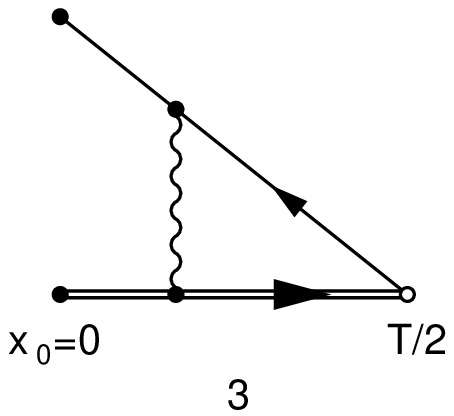,width=\linewidth}
  \end{minipage}
  \end{center}
  \caption{One loop diagrams contributing to $f_{\rm A}^{\rm
      stat}(T/2)$.}
  \label{f_fastat}
\end{figure}

\begin{figure}
  \noindent
  \begin{center}
  \begin{minipage}[b]{.3\linewidth}
    \centering\epsfig{figure=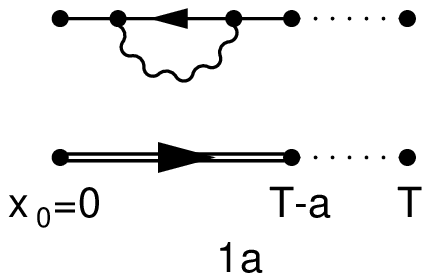,width=\linewidth}
  \end{minipage}
  \begin{minipage}[b]{.3\linewidth}
    \centering\epsfig{figure=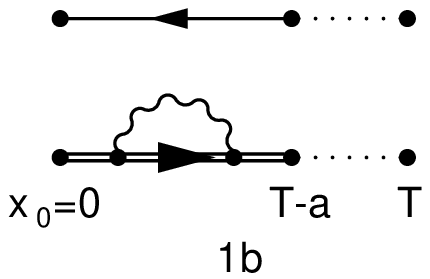,width=\linewidth}
  \end{minipage}
  \begin{minipage}[b]{.3\linewidth}
    \centering\epsfig{figure=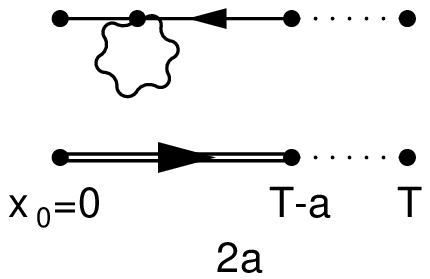,width=\linewidth}
  \end{minipage}\\
  \vspace{4mm}
  \begin{minipage}[b]{.3\linewidth}
    \centering\epsfig{figure=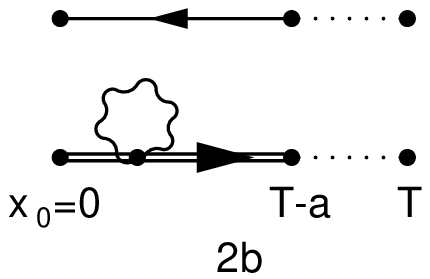,width=\linewidth}
  \end{minipage}
  \begin{minipage}[b]{.3\linewidth}
    \centering\epsfig{figure=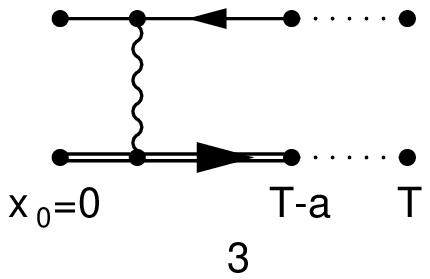,width=\linewidth}
  \end{minipage}
  \begin{minipage}[b]{.3\linewidth}
    \centering\epsfig{figure=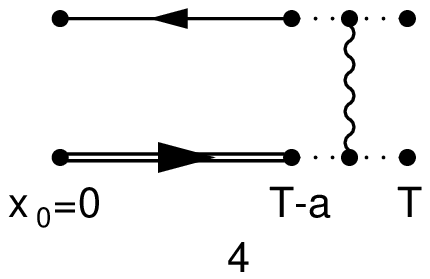,width=\linewidth}
  \end{minipage}\\
  \vspace{4mm}
  \begin{minipage}[b]{.3\linewidth}
    \centering\epsfig{figure=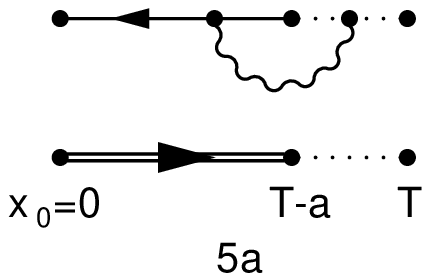,width=\linewidth}
  \end{minipage}
  \begin{minipage}[b]{.3\linewidth}
    \centering\epsfig{figure=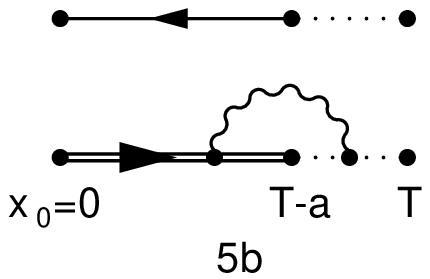,width=\linewidth}
  \end{minipage}
  \begin{minipage}[b]{.3\linewidth}
    \centering\epsfig{figure=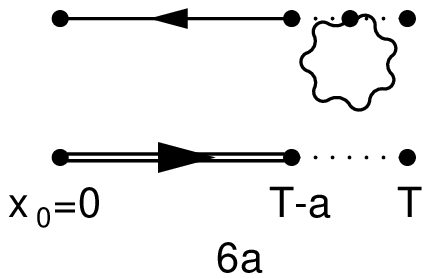,width=\linewidth}
  \end{minipage}\\
  \vspace{4mm}
  \begin{minipage}[b]{.3\linewidth}
    \centering\epsfig{figure=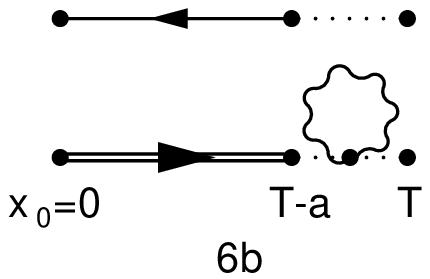,width=\linewidth}
  \end{minipage}
  \begin{minipage}[b]{.3\linewidth}
    \centering\epsfig{figure=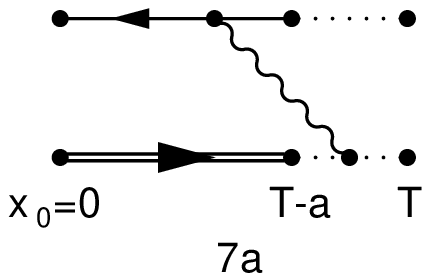,width=\linewidth}
  \end{minipage}
  \begin{minipage}[b]{.3\linewidth}
    \centering\epsfig{figure=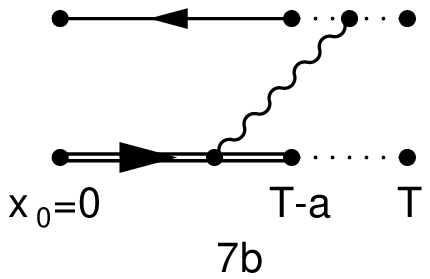,width=\linewidth}
  \end{minipage}
  \end{center}
  \caption{One loop diagrams contributing to $f_1^{\rm stat}$. The
    dotted lines denote the link from $T-a$ to T.}
  \label{f_f1stat}
\end{figure}

The classical equation of motion for the heavy quark,
\be
  \nabstar{0}\psi_{\rm h,cl}(x)=0,
\ee
is solved by
\be
  \psi_{\rm h,cl}(x)=U(x-a\hat{0},0)^{-1}\ldots
  U(x-x_0\hat{0},0)^{-1}\rhoh(\vecx),
\ee
which leads to
\be
  \Hheavy(x)=U(x-a\hat{0},0)^{-1}\ldots U(x-x_0\hat{0},0)^{-1}P_{+}.
\ee
Expanding this in powers of the bare coupling, and writing
\be
  q_0(x)=L^{-3}\sum_{\vectp}e^{i\vectp\vecx}\tilde{q}_0(x_0,\vectp)
\ee
we get
\be
  \Hheavy^{(0)}(x)=P_{+},
\ee
\be
  \Hheavy^{(1)}(x)=-{a\over L^3}\sum_{\vectp}\sum_{u_0=a}^{T-a}
    e^{i\vectp\vecx}\tilde{S}_{\rm h}(x_0,u_0;\vectp)^{(0)}
    \tilde{q}_0(u_0-a;\vectp)P_{+},
\ee
and
\be
  \Hheavy^{(2)}(x)=\Hheavy^{(2)}(x)_1+\Hheavy^{(2)}(x)_2
\ee
with
\bes
  \Hheavy^{(2)}(x)_1 & = & {{a^2}\over{L^6}}\sum_{\vectq,\vectp}
    \sum_{u_0=a}^{T-a}\sum_{v_0=a}^{T-a}e^{i(\vectq+\vectp)\vecx}
    \tilde{S}_{\rm h}^{(0)}(x_0,u_0;\vectq+\vectp)\tilde{q}_0(u_0-a;\vectq) \\
    & & \times\tilde{S}_{\rm h}^{(0)}(u_0-a,v_0;\vectp)
    \tilde{q}_0(v_0-a;\vectp)P_{+}
\ees
and
\bes
  \Hheavy^{(2)}(x)_1 & = & {{a^2}\over{2L^6}}\sum_{\vectq,\vectp}
    \sum_{u_0=a}^{T-a}e^{i(\vectq+\vectp)\vecx}
    \tilde{S}_{\rm h}^{(0)}(x_0,u_0;\vectn)\tilde{q}_0(u_0-a;\vectq) \\
    & & \times\tilde{q}_0(v_0-a;\vectp)P_{+}.
\ees
Contracting the gauge field fluctuations $\tilde{q}$ then gives the
diagrams in~\fig{f_fastat} and~\fig{f_f1stat}.

\subsection{Calculation of $(\castat)^{(1)}$ and $(\bastat)^{(1)}$
  \label{a_StatImpr}}

For the calculation of $(\castat)^{(1)}$, the one-loop coefficient of
the ratio $X_{\rm I}$ is decomposed as
\be
  X_{\rm I}^{(1)}(L/a)=
  X^{(1)}(L/a)|_{\cttilde=1}+\cttilde^{(1)}aX_{\rm b}^{(0)}(L/a)
  +{\castat}^{(1)}aX_{\delta\rm A}^{(0)}.
\ee
Defining the derivatives $\partial$ and $\partial^{\ast}$ by
\bes
  \partial f(L/a) & = & {1\over a}
    \left\{f\left(\frac{L+a}{a}\right)-f(L/a)\right\} \\
  \partial^{\ast} f(L/a) & = & {1\over a}
    \left\{f(L/a)-f\left(\frac{L-a}{a}\right)\right\},
\ees
the $\Oa$ part of $X_{\rm lat}^{(1)}$ can be extracted to get
\be
  {\castat}^{(1)}={{\lim_{a/L\rightarrow 0}
    \frac{L^2}{2a}(\partial+\partial^{\ast})
    X_{\rm lat}^{(1)}(\frac{L}{a})|_{\cttilde=1}
    -\lim_{a/L\rightarrow 0}
    \cttilde^{(1)}LX_{\rm b}^{(1)}(\frac{L}{a})}\over
    {\lim_{a/L\rightarrow 0}
    LX_{\delta\rm A}^{(0)}(\frac{L}{a})}}
\ee
with the light quark mass set to zero.

In a similar way, $(\bastat)^{(1)}$ can be calculated,
\bes
  {\bastat}^{(1)} & = & \left\{\lim_{a/L\rightarrow 0}
    \frac{L^2}{2a}(\partial+\partial^{\ast})
    \tilde{X}^{(1)}(\frac{L}{a})|_{\cttilde=1}
    -\lim_{a/L\rightarrow 0}
    \cttilde^{(1)}LX_{\rm b}^{(1)}(\frac{L}{a})\right.\nonumber\\
    && \left.\left.-\lim_{a/L\rightarrow 0}
    {\castat}^{(1)}LX_{\delta\rm A}^{(0)}(\frac{L}{a})\right\}\right/
    \left\{\lim_{a/L\rightarrow 0}zX^{(0)}(\frac{L}{a})\right\},
  \label{e_BastatForm}
\ees  
where
\be
  \tilde{X}^{(1)}=(1+{\bastat}^{(0)}am_{\rm q}^{(0)})X^{(1)}
    +(Z_{\rm A, lat}^{\rm stat (1)}+
    {\bastat}^{(0)}am_{\rm q}^{(1)})X^{(0)}.
\ee
The quantity
\be
  z=Z_{\rm m, lat}(1+b_{\rm m}am_{\rm q})Lm_{\rm q}
\ee
is kept fixed when taking the limit. The renormalization and
improvement constants for the light quark mass are known
from~\cite{impr:pap5}.

%% file: appc.tex
\section{Numerical results
  \label{a_Numeric}}

The following results were obtained at $T=L$,
with the bare light quark mass set 
to the critical mass
\be
  m_{\rm c}=m_{\rm c}^{(1)}g_0^2+O(g_0^4),
\ee
with~\cite{impr:pap2}
\be
  am_{\rm c}^{(1)}=-0.2025565\times C_{\rm F}.
\ee
The improvement coefficients used are
\be
  \csw=1,
\ee
and
\be
  \cttilde=1+\cttilde^{(1)}g_0^2+O(g_0^4),
\ee
with~\cite{impr:pap5}
\be
  \cttilde^{(1)}=-0.01346\times C_{\rm F}.
\ee

The tree level results can be obtained analytically. Using the
notation of~\cite{impr:pap2}, they are
\bes
  {\fastat}^{(0)}(x_0) & = & -{3\over{R(p^{+})}}
    \{(M(p^{+})-\stackrel{\circ}{p}_0^{+})
    e^{-\omega(\vectp^{+})x_0} \nonumber\\
  & &-((M(p^{+})+\stackrel{\circ}{p}_0^{+})
    e^{-\omega(\vectp^{+})(2L-x_0)}\}\qquad{\rm at}\quad\vectp=\vectn,
\ees
and
\bes
  {\fonestat}^{(0)} & = & {3\over{R(p^{+})}}
    \{(M(p^{+})-\stackrel{\circ}{p}_0^{+})
    e^{-\omega(\vectp^{+})(L-a)} \nonumber\\
    & & -((M(p^{+})+\stackrel{\circ}{p}_0^{+})
    e^{-\omega(\vectp^{+})(L+a)}\}\qquad{\rm at}\quad\vectp=\vectn.
\ees
Also the axial current improvement term can be calculated analytically 
at tree level,
\be
  {\fdeltaastat}^{(0)}(x_0)=
    -{3(\stackrel{\circ}{\vectp}^{+})^2\over{R(p^{+})}}
    \{e^{-\omega(\vectp^{+})x_0}-e^{-\omega(\vectp^{+})(2L-x_0)}\}
    \qquad{\rm at}\quad\vectp=\vectn.
\ee
The one loop results are given tables~(\ref{t_fxstat1_t0.0})--
(\ref{t_fxstat1_t1.0}). They have been
obtained in double precision calculations. Where an error is given, it 
has been taken from the comparison with the corresponding quadruple
precision result.

\input{fx1_t0.0}

\input{fx1_t0.5}

\input{fx1_t1.0}

\clearpage

%% file: fx1_t0.0.tex
\begin{table}[p]
\centering
\begin{tabular}{r r@{.}l r@{.}l}
\hline \\[-1.0ex]
\multicolumn{1}{c}{$L/a$} & \multicolumn{2}{c}{${\fastat}^{(1)}(L/2)$} &
\multicolumn{2}{c}{${\fonestat}^{(1)}$} \\[1.0ex] \hline \\[-1.0ex]
  4 &   1&1696384959451731(7) & $-2$&788717306605536(2) \\
  6 &   1&5721721053841535 & $-3$&744847208299456 \\
  8 &   2&049414037209983(4) & $-4$&76922384055426(4) \\
 10 &   2&543921176114074 & $-5$&80311014058304 \\
 12 &   3&04401042107373(4) & $-6$&8365479914505(2) \\
 14 &   3&54641935327602 & $-7$&86798061045362 \\
 16 &   4&04997105269991(6) & $-8$&897379383110(2) \\
 18 &   4&55415464879211 & $-9$&924989104618 \\
 20 &   5&058717578768(1) & $-10$&951072803831(9) \\
 22 &   5&5635226305105 & $-11$&975859714707 \\
 24 &   6&0684897524216 & $-12$&999539542806 \\
 26 &   6&5735695645449 & $-14$&02226731358 \\
 28 &   7&0787302119581 & $-15$&04416980198 \\
 30 &   7&583950374904 & $-16$&06535128339 \\
 32 &   8&089215337913 & $-17$&08589819107 \\
 34 &   8&594514673902 & $-18$&10588276384 \\
 36 &   9&099840825103 & $-19$&12536587521 \\
 38 &   9&60518820413 & $-20$&14439922726 \\
 40 &  10&11055260756 & $-21$&1630270594 \\
 42 &  10&61593082380 & $-22$&1812874905 \\
 44 &  11&12132036473 & $-23$&1992135777 \\
 46 &  11&62671927843 & $-24$&2168341613 \\
 48 &  12&1321260162(3) & $-25$&234174544(1) \\[1.0ex]
\hline\\
\end{tabular}
\caption{\footnotesize $\fastat$ and $\fonestat$ at one loop order with
$\theta=0.0$
\label{t_fxstat1_t0.0}}
\end{table}

%% file: fx1_t0.5.tex
\begin{table}[p]
\centering
\begin{tabular}{r l r@{.}l}
\hline \\[-1.0ex]
\multicolumn{1}{c}{$L/a$} & \multicolumn{1}{c}{${\fastat}^{(1)}(L/2)$} &
\multicolumn{2}{c}{${\fonestat}^{(1)}$} \\[1.0ex] \hline \\[-1.0ex]
  4 &   0.707462157465912(1) & $-1$&828520866963525(4) \\
  6 &   1.0665692749675324 & $-2$&53494068630446456 \\
  8 &   1.45264489740883(2) & $-3$&268670455891038(1) \\
 10 &   1.845191095078847 & $-4$&0063158459818604 \\
 12 &   2.23966783029017(2) & $-4$&7434246150126(1) \\
 14 &   2.634843388684698 & $-5$&47929315793510 \\
 16 &   3.03031175469344(4) & $-6$&2139336444356(4) \\
 18 &   3.42591623855986 & $-6$&9474888408264 \\
 20 &   3.82158857585887 & $-7$&6801091376222 \\
 22 &   4.21729615057110 & $-8$&4119266083201 \\
 24 &   4.61302225619736 & $-9$&1430518919882 \\
 26 &   5.00875788592884 & $-9$&8735765450344 \\
 28 &   5.40449800734282 & $-10$&603576331834 \\
 30 &   5.80023974629234 & $-11$&333114250324 \\
 32 &   6.19598144778345 & $-12$&062243035152 \\
 34 &   6.59172216643778 & $-12$&791007154176 \\
 36 &   6.98746137900958 & $-13$&51944438137 \\
 38 &   7.3831988169726 & $-14$&24758703382 \\
 40 &   7.7789343665427 & $-14$&97546294777 \\
 42 &   8.1746680077261 & $-15$&70309625284 \\
 44 &   8.5703997766495 & $-16$&43050799027 \\
 46 &   8.9661297419670 & $-17$&1577166104 \\
 48 &   9.36185799001(1) & $-17$&8847383757(5) \\[1.0ex]
\hline\\
\end{tabular}
\caption{\footnotesize $\fastat$ and $\fonestat$ at one loop order with
$\theta=0.5$
\label{t_fxstat1_t0.5}}
\end{table}

%% file: fx1_t1.0.tex
\begin{table}[p]
\centering
\begin{tabular}{r l l}
\hline \\[-1.0ex]
\multicolumn{1}{c}{$L/a$} & \multicolumn{1}{c}{${\fastat}^{(1)}(L/2)$} &
\multicolumn{1}{c}{${\fonestat}^{(1)}$} \\[1.0ex] \hline \\[-1.0ex]
  4 &   0.3044333805556540(7) & $-0$.869671412347571(1) \\
  6 &   0.5486322003500290 & $-1$.1948835905593809 \\
  8 &   0.793695141003886(4) & $-1$.53524521245116(1) \\
 10 &   1.038456376564977 & $-1$.881017287644124 \\
 12 &   1.28269652915735(3) & $-2$.22894706469196(3) \\
 14 &   1.52651219643810 & $-2$.57783369422662 \\
 16 &   1.77002559881285(11) & $-2$.92715777151496(7) \\
 18 &   2.01332870498098 & $-3$.27666836444461 \\
 20 &   2.2564839354402(4) & $-3$.6262359702708(2) \\
 22 &   2.4995329275626 & $-3$.97579139115074 \\
 24 &   2.7425037517644 & $-4$.3252973303331 \\
 26 &   2.9854157215170 & $-4$.6747341210386 \\
 28 &   3.2282824349472 & $-5$.0240921351849 \\
 30 &   3.4711136865628 & $-5$.3733675663630 \\
 32 &   3.7139166829809 & $-5$.7225600072263 \\
 34 &   3.9566968307148 & $-6$.071671021232 \\
 36 &   4.199458257584 & $-6$.420703284271 \\
 38 &   4.442204165276 & $-6$.769660062007 \\
 40 &   4.684937072624 & $-7$.118544889597 \\
 42 &   4.927658986791 & $-7$.467361375608 \\
 44 &   5.170371525823 & $-7$.81611308330 \\
 46 &   5.413076007854 & $-8$.16480346057 \\
 48 &   5.65577351702(2) & $-8$.51343580067(7) \\[1.0ex]
\hline\\
\end{tabular}
\caption{\footnotesize $\fastat$ and $\fonestat$ at one loop order with
$\theta=1.0$
\label{t_fxstat1_t1.0}}
\end{table}